\documentclass[sigconf,letterr]{acmart}

\usepackage[absolute,showboxes]{textpos}

\usepackage[utf8]{inputenc}
\usepackage{booktabs}
\usepackage{subcaption}
\usepackage[inline]{enumitem}
\usepackage{balance}
\usepackage{stfloats}
\usepackage{standalone}
\usepackage{pgfplots}
\usepackage{tikz}
\usetikzlibrary{arrows.meta,calc,colorbrewer,external,patterns,shapes,pgfplots.statistics}
\usepgfplotslibrary{groupplots,colormaps,colorbrewer}
\usepackage{pgfplotstable}
\pgfplotsset{compat=1.14}

\usepackage{xspace}

\newcommand{\one}{({\em i})\xspace}
\newcommand{\two}{({\em ii})\xspace}

\acmYear{2019}\copyrightyear{2019}
\setcopyright{none}
\acmConference[ICN '19]{ICN '19: Conference on Information-Centric Networking}{September 24--26, 2019}{Macau, Macao}
\acmBooktitle{ICN '19: Conference on Information-Centric Networking, September 24--26, 2019, Macau, Macao}
\acmPrice{15.00}
\acmDOI{10.1145/3357150.3357404}
\acmISBN{978-1-4503-6970-1/19/09}

\renewcommand\footnotetextcopyrightpermission[1]{}

\graphicspath{{figs/}}
\tikzexternaldisable

\begin{document}
\title[QoS Management in Constrained NDN Networks]{Gain More for Less: The Surprising Benefits of \\  QoS Management in Constrained NDN Networks}

\author{Cenk G{\"u}ndo\u{g}an}
\affiliation{%
  \institution{HAW Hamburg}
}
\email{cenk.guendogan@haw-hamburg.de}

\author{Jakob Pfender}
\affiliation{%
  \institution{Victoria University of Wellington}
}
\email{jpfender@ecs.vuw.ac.nz}

\author{Michael Frey}
\affiliation{%
  \institution{Safety IO}
}
\email{michael.frey@safetyio.com}

\author{Thomas C. Schmidt}
\affiliation{%
  \institution{HAW Hamburg}
}
\email{t.schmidt@haw-hamburg.de}

\author{Felix Shzu-Juraschek}
\affiliation{%
  \institution{Safety IO}
}
\email{felix.juraschek@safetyio.com}

\author{Matthias W{\"a}hlisch}
\affiliation{%
  \institution{Freie Universit\"at Berlin}
}
\email{m.waehlisch@fu-berlin.de}

\renewcommand{\shortauthors}{C. G{\"u}ndo\u{g}an et al.}

\begin{abstract}
  Quality of Service (QoS) in the IP world mainly manages forwarding resources, i.e., link capacities and buffer spaces. In addition, Information Centric Networking (ICN) offers  resource dimensions such as in-network caches and forwarding state. In  constrained wireless networks, these resources  are scarce  with a potentially high impact due to lossy radio transmission. In this paper, we explore the two basic service qualities (i) \emph{prompt} and (ii) \emph{reliable} traffic forwarding for the case of NDN. The resources we take into account are forwarding and queuing priorities, as well as the utilization of caches and of forwarding state space. We treat QoS resources not only in isolation, but correlate their use on local nodes and between network members. Network-wide coordination is based on simple, predefined QoS code points. Our findings indicate that coordinated QoS management in ICN is more than the sum of its parts and exceeds the impact QoS can have in the IP world.
\end{abstract}


\maketitle

\setlength{\TPHorizModule}{\paperwidth}
\setlength{\TPVertModule}{\paperheight}
\TPMargin{5pt}
\begin{textblock}{0.8}(0.1,0.02)
     \noindent
     \footnotesize
     If you cite this paper, please use the ICN reference:
     C. G{\"u}ndo\u{g}an, J. Pfender, M. Frey, T.~C. Schmidt, F. Juraschek, M. W\"ahlisch. Gain More for Less: The Surprising Benefits of QoS Management in Constrained NDN Networks. In \emph{Proc. of ACM ICN}, ACM, 2019.
\end{textblock}

\section{Introduction}\label{sec:intro}

	Quality of Service (QoS) in IP networks has been around for two decades, but so far has experienced remarkably little deployment. Its hesitant adoption is commonly understood to have two reasons: limited scalability  (IntServ~\cite{RFC-1633,RFC-2205}) and plain resource trading (DiffServ~\cite{RFC-2474,RFC-2475})---the latter is often referred to as managed unfairness. While QoS in the IP world is mainly restricted to managing forwarding resources (link capacities and buffer spaces)~\cite{w-nccmi-05} and the IEEE follows this approach on Layer~2 with Time Sensitive Networking (TSN)~\cite{ieee8021q-18,msks-eatts-13}, Information-Centric Networking (ICN)~\cite{adiko-sind-12,xvsft-sinr-14}  offers additional resource dimensions such as in-network caches and forwarding states that can shape network performance significantly.

Constrained wireless networks used to be the playground for experimental research and tinkering, but have recently turned into a fast-growing market. The advent of the Internet of Things (IoT) introduced the  vision of omnipresent and always connected sensors and actuators that generate  business models from new products, innovative processes, and data---a total of 1.6 Zettabytes is soon expected~\cite{ms-eai-15} in this rapidly growing business segment. LoRa~\cite{la-toll-15} and NB-IoT~\cite{nb-iot-16} have already started deployment for a wide-area outreach to the embedded edge. As of today, 5G technologies have appeared on the horizon with the promise  of tailored technologies that can deliver dedicated service qualities to their users in vertically sliced networks.

Key use cases for assured QoS are raised by industry. The Industrial Internet-of-Things (IIoT) is on track to interconnect time- and safety-critical infrastructure in industrial environments consisting of embedded microcontrollers with nearby edge intelligent and  remote cloud services. Its use cases continually require high reliability and very low latency. 
Traditionally, these sensors and actuators are wired via  field buses, which have been the state-of-the-art for decades, but often lack interconnectivity, interoperability, and security.
Communication infrastructure components in industrial environments are also often physically separated from one another.

5G facilitates the creation of networks that tackle these challenges and allows companies to create their own private 5G-based networks on site.
Companies can provide their own 5G infrastructure  by making use of a key 5G concept called network slicing. 
Network slicing enables the creation of sub-networks for specific services and users that can preselect 5G network parameters such as end-to-end latency, maximum throughput, and traffic density.
It allows companies to deploy ultra-reliable low-latency networks for critical infrastructure.

ICN has been a promising candidate for networking the IoT edge for a while~\cite{bmhsw-icnie-14,pf-britu-15,acim-icnis-15,sblwy-ndnti-16,arxpp-kip-17,szsmb-avdir-17}. Recent experimental studies~\cite{gklp-ncmcm-18,caamr-uisfb-18} confirmed that ICN can sustain high reliability at moderate latency penalty   even on lossy wireless links, and showed how it  could be integrated into the 5G edge architecture. Nevertheless, current analyses and solutions are built purely on equal resource sharing.  Studies of ICN behavior under prioritization and active QoS management are missing.

In this paper, we perform a first exploration of QoS management impacts on NDN~\cite{jstp-nnc-09,zabjc-ndn-14} in the resource-constrained IoT using RIOT~\cite{bghkl-rosos-18}. We start from the two basic service attributes {\em (i) prompt} and {\em (ii) reliable} traffic forwarding and define a simple yet efficient management scheme. We carefully implement these QoS semantics by employing not only the NDN resources \emph{forwarding capacity, PIT state,} and \emph{cache} alone, but correlating resources internally on a node and also externally between nodes. These correlations can be performed without additional signaling overhead.

Our findings from extensive experiments in a large testbed confirm the efficacy of our approach. To our surprise, however, the QoS measures do not sacrifice the performance of unprioritized traffic. Even though resources shift to the prime packets, best effort  flows still uphold their performance or even improve. A thorough analysis reveals the positive effects of resource correlation, which raises the overall network performance to a higher level than in the state of uncoordinated  resource allocation.

The remainder of this paper is structured as follows. The problem space of managing distributed resources is discussed in Section~\ref{sec:related-work} along with related work. Section~\ref{sec:icn-qos} introduces the QoS building blocks and how we manage the resources. Our implementations and evaluations are presented in Section~\ref{sec:eval}, followed by the conclusions and an outlook on future work in Section~\ref{sec:c+o}.

\section{The Problem of Distributed Resource Management in ICN}\label{sec:related-work}

\subsection{Problem Statement}

Implementing service differentiation and assurance in a network raises the challenge of managing distributed resources without sacrificing them. Common Internet approaches follow a flow-based (e.g., IntServ) or a class-based (e.g., DiffServ) concept. Flow-based resource reservation requires dedicated signaling and state, which quickly reaches scalability limits with resource exhaustion. In the presence of ubiquitous in-network caching and request aggregation, content endpoints are unspecified and data paths are inherently multi-source, multi-destination, and possibly widely disjoint. This makes ICN flows difficult to identify and to maintain.

\begin{figure}
  \centering
  \tikzset{
  pit/.style = { minimum width=1.75cm,minimum height=1.25cm,draw,rounded corners,label={[anchor=north,font=\bfseries,yshift=-0.1cm]above:PIT}, fill=Pastel2-F, inner sep=0pt},
  pite/.style = { rounded corners=1pt,minimum width=0.3cm, minimum height=0.4cm, inner sep=0pt, anchor=south, draw=black },
  piteleft/.style = { pite, yshift=0.15cm, xshift=-0.45cm },
  pitemid/.style = { pite, yshift=0.15cm },
  piteright/.style = { pite, yshift=0.15cm, xshift=0.45cm },
  pite0/.style = { piteleft, fill=Pastel2-B },
  pite1/.style = { pitemid, fill=Pastel2-B },
  pite2/.style = { piteright, fill=Pastel2-F },
  cs/.style = {draw,cylinder,minimum width=1.25cm,minimum height=0.70cm,shape border rotate=90,inner sep=3pt,cylinder uses custom fill, cylinder body fill=Pastel2-F,cylinder end fill=Pastel2-F!50},
  queue/.style = { draw, cylinder, minimum width=0.75cm,minimum height=2.15cm,inner sep=3pt,cylinder uses custom fill, cylinder body fill=Pastel2-F, cylinder end fill=Pastel2-F!50 },
  packet/.style = { queue, minimum width=0.5cm,minimum height=0.4cm,inner sep=3pt,cylinder uses custom fill, cylinder body fill=Pastel2-B, cylinder end fill=Pastel2-B!50,anchor=west },
}

\begin{tikzpicture}[>={Latex}]
  \node[queue] (q0) at (0.0cm,0.5cm) {};
  \node[queue,label={[anchor=north,font=\bfseries,xshift=-0.45cm,yshift=-0.05cm]below:{Forwarding Queues}}] (q1) at (0.0cm,-0.5cm) {};
  \node[packet] (p1) at ([xshift=0.2cm]q0.west) {};
  \node[packet] (p2) at ([xshift=0.35cm]p1.west) {};
  \node[packet] (pd1) at ([xshift=0.2cm]q1.west) {};
  \node[packet] (pd2) at ([xshift=0.35cm]pd1.west) {};
  \node[packet] (pd3) at ([xshift=0.35cm]pd2.west) {};
  \node[packet,anchor=west] (p7) at ([xshift=-1.2cm]$(q0.west)!0.5!(q1.west)$) {};
  \node[packet,anchor=west] (p8) at ([xshift=-0.1cm]p7.east) {};
  \draw[->] (p8) -- (q0);
  \draw[->] (p8) -- (q1);

  \node[pit,anchor=west] (pit0) at ([xshift=0.8cm]$(q0.east)!0.5!(q1.east)$) {};
  \node[pite0] at (pit0.south) {};
  \node[pite1] at (pit0.south) {};
  \node[pite2] at (pit0.south) {};

  \node[cs,cylinder body fill=Pastel2-B,cylinder end fill=Pastel2-B!50,font=\bfseries\small,anchor=west] (cs0) at ([xshift=0.8cm,yshift=-0.5cm]pit0.east) {};
  \node[cs,anchor=south,yshift=-0.150cm,cylinder body fill=Pastel2-B,cylinder end fill=Pastel2-B!50] (cs1) at (cs0.north) {};
  \node[cs,anchor=south,yshift=-0.150cm] (cs2) at (cs1.north) {};
  \node[font=\bfseries] at (cs2) {CS};

  \draw[thick] ([xshift=-0.2cm,yshift=1.0cm]p7.west) coordinate (a) -- ++(0cm,0.3cm) coordinate (b) -- node[pos=0.5,fill=white] {IP Resources} ([xshift=0.2cm]b -| q0.east) -- ++(0cm,-0.3cm);
  \draw[thick] ([xshift=-0.2cm,yshift=-1.3cm]p7.west) coordinate (a) -- ++(0cm,-0.3cm) coordinate (b) -- node[pos=0.5,fill=white] {NDN Resources} ([xshift=0.2cm]b -| cs0.east) -- ++(0cm,0.3cm);
\end{tikzpicture}%
  \caption{Manageable resources in IP vs. NDN.}
  \label{fig:resources}
\end{figure}
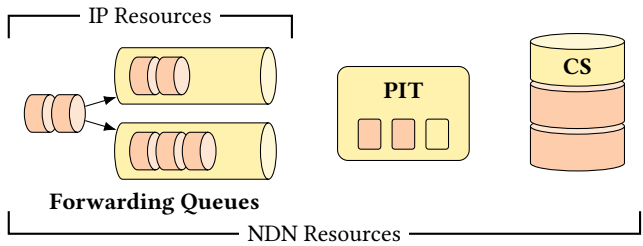

Resource allocation according to packet classes requires ingress shaping and filtering, since unforeseen traffic bursts quickly exhaust the per-class reservations and counteract service assurances. In several ICN flavors including NDN and CCNx, link occupancy and forwarding demands are steered hop-by-hop in a request-response fashion. Small requests trigger data replies of unknown size, provenance, and timing. This complicates reliable resource predictions for reponses in NDN.\@ Shaping and dropping Interests can prevent resource exhaustion, but may leave the network underutilized. Restricting ingress only to data may lead to bursts of unsatisfied Interests, which waste network resources.

In-network caches in ICN enrich the field of manageable resources. Caches reduce latency and forwarding load and often take the role of a (large, delay-tolerant) retransmission buffer. With NDN/CCNx, additional resources come into play in the form of Pending Interest Tables (PITs) that govern stateful forwarding. The overall resource ensemble is visualized in Figure~\ref{fig:resources} and raises concerns. Capacities in forwarding, caching, and pending Interest state may be largely heterogeneous. A wirespeed forwarder may supply negligible cache memory  compared to its transmission capacity, for example. In the IoT the opposite is often true in that flash memory is normally shipped in `infinite' sizes when compared to the main memory (PIT) or the wireless data rate. A beneficial resource management faces the problem of how to carefully balance these resources and arrive at an overall optimized network performance~\cite{RFC-7927}.

\begin{figure}
  \centering
  \tikzset{
  pit/.style = { minimum width=1.75cm,minimum height=1.25cm,draw,rounded corners,label={[anchor=north,font=\bfseries,yshift=-0.1cm]above:PIT}, fill=Pastel2-F },
  pite/.style = { rounded corners=1pt,minimum width=0.3cm, minimum height=0.4cm, inner sep=0pt, anchor=south, draw=black },
  piteleft/.style = { pite, yshift=0.2cm, xshift=-0.40cm },
  pitemid/.style = { pite, yshift=0.2cm },
  piteright/.style = { pite, yshift=0.2cm, xshift=0.40cm },
  pite0/.style = { piteleft, fill=Pastel1-A },
  pite1/.style = { pitemid, fill=Pastel1-B, postaction={ pattern=crosshatch } },
  pite2/.style = { piteright, fill=Pastel1-C, postaction={ pattern=north west lines } },
  pite3/.style = { piteleft, fill=Pastel1-B, postaction={ pattern=crosshatch } },
  pite4/.style = { pitemid, fill=Pastel1-D, postaction={ pattern=north east lines } },
  pite5/.style = { piteright, fill=Pastel1-E, postaction={ pattern=crosshatch dots } },
  pite6/.style = { piteleft, fill=Pastel1-A },
  pite7/.style = { pitemid, fill=Pastel1-D, postaction={ pattern=north east lines } },
  pite8/.style = { piteright, fill=Pastel1-E, postaction={ pattern=crosshatch dots } },
}

\begin{tikzpicture}[>={Latex}]
  \node[pit] (pit0) at (-3.30cm,0) {};
  \node[pit] (pit1) at (0,0) {};
  \node[pit] (pit2) at (3.30cm,0) {};
  \node[pite0] at (pit0.south) {};
  \node[pite1] at (pit0.south) {};
  \node[pite2] at (pit0.south) {};
  \node[pite3] at (pit1.south) {};
  \node[pite4] at (pit1.south) {};
  \node[pite5] at (pit1.south) {};
  \node[pite6] (s1) at (pit2.south) {};
  \node[pite7] (s2) at (pit2.south) {};
  \node[pite8] at (pit2.south) {};

  \draw[-] ([yshift=0.3cm]pit2.west) -- node[pos=1,font=\bfseries\ttfamily\large,inner sep=0pt] {x} ([yshift=0.3cm,xshift=-0.15cm]pit1.east);
  \path ([yshift=0.3cm]pit2.west) -- node[pite,pos=0.5,anchor=center,fill=Pastel1-A] {} ([yshift=0.3cm]pit1.east);
  \draw[->] ([yshift=-0.3cm]pit2.west) -- node[pite,pos=0.5,anchor=center,fill=Pastel1-D,postaction={ pattern=north east lines}] {} ([yshift=-0.3cm]pit1.east);
  \draw[-] ([yshift=-0.3cm]pit1.west) -- node[pos=1,font=\bfseries\ttfamily\large,inner sep=0pt] {x} ([yshift=-0.3cm,xshift=-0.15cm]pit0.east);
  \path ([yshift=-0.3cm]pit1.west) -- node[pite,pos=0.5,anchor=center,fill=Pastel1-D,postaction={ pattern=north east lines}] {} ([yshift=-0.3cm]pit0.east);
\end{tikzpicture}%
  \caption{PIT decorrelation terminates data paths.}
  \label{fig:pitdecorrelation}
\end{figure}
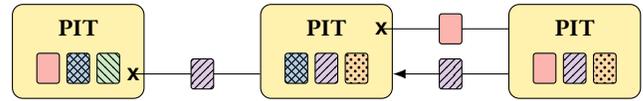

Resource complexity, however, extends beyond a single system. The impact  of distributed resources is easily flawed if management cannot  jointly coordinate contributions. Neighboring caches, for instance, are less effective if filled with identical data. A more delicate problem arises from PIT state management. If neighboring PITs  diverge and no longer represent common forwarding paths (see Figure~\ref{fig:pitdecorrelation}) all data flows terminate and forwarding resources are wasted. This problem of state decorrelation was first reported in~\cite{wsv-bdpts-13}.

\subsection{QoS Qualifiers}

QoS extensions for ICN have recently attracted attention and generated various efforts within the IRTF ICN research group~\cite{draft-oran-icnrg-qosarch,draft-oran-icnrg-flowbalance,draft-moiseenko-icnrg-flowclass,draft-anilj-icnrg-dnc-qos-icn,draft-icnrg-iotqos}.
A proposed flow classification mechanism~\cite{draft-moiseenko-icnrg-flowclass} differentiates traffic flows based on content name prefixes using two different methods.
The first method, \textit{EC3}, introduces a new message header entry in Data packets that indicates the prefix length of the
content name characteristic for  the classification process.
Once a consumer learns about such an equivalence class, it can also include equivalence class indicators into subsequent Interests.
The second method, \textit{ECNCT}, encodes classification indicators directly into names at content creation using a new type of name component.
This solution does not inflate messages with additional headers, which is advantageous for constrained IoT deployments.
Nevertheless, encoding flow classification indicators in typed name components leads to an inflation of names in the routing system. Identical content published with different classification values will lead to duplicate names that cannot be aggregated in  Forwarding Information Bases (FIBs). 
Analogously, the default matching functions used for the Pending Interest Table (PIT) and Content Store (CS)
will consider names that differ only in their flow classification as dissimilar, which conflicts with Interest aggregation and cache utility.

In the Internet Research Task Force (IRTF), two approaches to handle QoS treatments are under discussion.
An end-to-end QoS framework~\cite{draft-anilj-icnrg-dnc-qos-icn}, in which non-routable QoS markers
are appended as  suffixes to content names.
In analogy to DiffServ, these markers are then used to apply different resource allocation mechanisms to the request and response messages.
The FIB is extended to ignore QoS markers, and the PIT is modified to disaggregate pending requests that have differing QoS markers.
This disaggregation may lead to PIT inflation in particular for  setups with a high diversity of QoS markers.
The work at hand has also been presented to the IRTF~\cite{draft-icnrg-iotqos}.

Tsilopoulos~\emph{et~al.}~\cite{tx-sdtti-11} identify three different types of traffic based on two characteristics in ICN traffic.
The authors introduce two extensions to CCN in order to handle these traffic types, \emph{Persistent Interests} and \emph{Reliable Notifications}.
Persistent Interests are valid for Data packets which are produced during a pre-defined period of time.
Reliable Notifications inform receivers that real-time data is available and are propagated reliably on a hop-by-hop basis.
Notifications that are not acknowledged in time are retransmitted. 
If a receiver  successfully receives a notification but does not receive data in a given time, new Interests  are created to renew the request.

\subsection{Distributed Forwarding Resources}

MIRCC~\cite{mago-mmirc-16} introduces a rate-based, multipath-aware congestion control scheme for ICN.\@
Each Data message in MIRCC contains a rate value which in turn is used to calculate per-link rates.
It is inspired by the Rate Control Protocol (RCP)~\cite{d-rcpcc-08} for IP networks.

Another approach~\cite{nbvr-qsin-14} manages forwarding resources and is experimentally evaluated for the PURSUIT architecture.
This work implements a QoS differentiation scheme, where each forwarder manages virtual links that include
packet queues with varying traffic rates and a designated traffic shaper.
QoS information is encoded into the names and determines the mapping of traffic flows to low or high priority virtual links.

\subsection{Distributed Cache Management}%
\label{subsec:relwork_caching}

Caching policies that employ heuristics to inform a caching decision
instead of caching all incoming content can be broadly organized into
a number of different families, depending on what information they use
to reach their caching decision.

The easiest way to achieve higher cache diversity without
increasing the complexity of the caching policy is to cache
probabilistically. The static version of this approach, commonly
known as $Prob(p)$, uses a static probability $p$ that governs how
likely a given node will cache a given content chunk. It has been
shown~\cite{zhang_survey_2015, hail_caching_2015} that $Prob(p)$
outperforms the default strategy of caching everything in terms of cache
diversity, and that lower values for $p$ correlate with higher
diversity~\cite{psaras_probabilistic_2012, tarnoi_performance_2014,
hail_performance_2015, hail_caching_2015, zhang_survey_2015,
arshad_information-centric_2017}.

Instead of using the same static caching probability for all incoming
content, a caching strategy may also dynamically compute a probability
for each individual node or even for each content chunk in order to
adapt the caching behavior to the state of the network. These
strategies may be based purely on node-local information, such as CS
saturation or battery levels; on properties of the incoming content,
such as its name, freshness, type, or producer; or on information from the
wider network, such as the position of the caching node in the network
topology or the CS contents of neighboring nodes. Examples include
\emph{ProbCache}~\cite{psaras_probabilistic_2012}, which computes the
caching probability of a given content chunk based on the total number
of hops between its producer and the consumer that requested it, and
\emph{pCASTING}~\cite{hail_caching_2015}, which considers the freshness
of the content as well as the node's battery level and CS saturation
when calculating $p$. Both strategies have been found to increase the
cache hit ratio, reduce the average number of hops required to hit
requested content, and reduce the number of cache
evictions~\cite{pfender_performance_2018}. Various other dynamic
probabilistic caching strategies have been proposed, with decisions
based on content freshness~\cite{hail_caching_2015,
doan_van_efficient_2018}, content popularity~\cite{chen_brr-cvr:_2016}, or
whether the content is already in a neighboring
CS~\cite{zhang_lf:_2015}.

Not all caching strategies use the probabilistic approach. Instead, some
exploit knowledge about the network topology~\cite{chai_cache_2013,
naz_dynamic_2018, zhou_broadcasting_2015, sun_topology_2016, iy-jorca-17}, which has
the advantage of taking global knowledge about the network into account
but often comes at significant costs such as lengthy setup times,
communications and memory overhead, and vulnerability to changes in the
topology~\cite{pfender_content_2019}. Another class of caching policies
has nodes cooperate with their neighbors, either explicitly by
exchanging information~\cite{liu_novel_2016} or implicitly by using
pre-defined rules~\cite{li_time-shifted_2011, zeng_caching_2011,hbswa-litnc-17}.


\section{QoS Building Blocks for NDN}\label{sec:icn-qos}

\subsection{Traffic Flow Classification}%
\label{subsec:flowclass}
General purpose networks simultaneously host competing traffic flows that exhibit varying resource requirements and time constraints.
This also holds for typical IoT deployments, in which flows originate from sensors and actuators, or from remote cloud services that connect via gateways to the IoT domain.
A flow classification is necessary for differentiating packets that belong to separate message flows, whenever the network should allocate distinct resources and treat them in a differentiated manner.

In the IP world, traffic flows are defined by the application endpoints and identified by address/port tuples (IPv4) or addresses plus flow label (IPv6). Since ICN abandons the host-centric paradigm, this definition of traffic flows  no longer holds. In the presence of delocalized content and in-network caching, the concept of application endpoints becomes meaningless. 

The characterizing property of these packets are content names (or prefixes). In the example of CCNx and NDN, content names appear  in related Interest and Data packets.
This ubiquity of names and their potential to impose hierarchies on content make them a distinguished component for identifying flows. Accordingly, we propose a traffic flow classification mechanism for NDN that builds on hierarchical names, prefixes, and longest common prefix match. 
It is explicitly designed to not put a strain on typically resource constrained
IoT devices. In this regard, our scheme is computationally simple and does not require an additional overhead in message headers.

\begin{figure}
  \centering
  \begin{tikzpicture}
  \draw[-{Latex[scale=2]}] (-0.5cm,0) -- node[below,sloped] {Latency} ++(8.5cm,0);
  \draw[-{Latex[scale=2]}] (0,-0.5cm) -- node[above,sloped,pos=0.55] {Reliability} ++(0,2.5cm);
  \node[font=\bfseries] at (2.15cm,0.5cm) {$\langle$ Regular, Prompt $\rangle$};
  \node[font=\bfseries] at (2.15cm,1.25cm) {$\langle$ Reliable, Prompt $\rangle$};
  \node[font=\bfseries] at (5.75cm,0.5cm) {$\langle$ Regular, Regular $\rangle$};
  \node[font=\bfseries] at (5.75cm,1.25cm) {$\langle$ Reliable, Regular $\rangle$};
\end{tikzpicture}%
  \caption{QoS Service Levels.}%
  \label{fig:qosdim}
\end{figure}
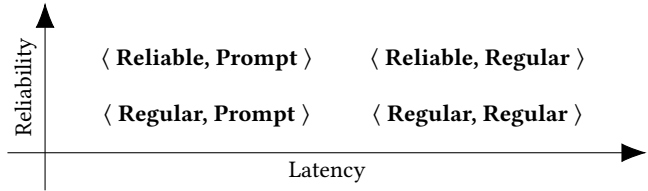

In this work, we consider service differentiation with respect to two quality dimensions: \emph{latency} and \emph{reliability}. For simplicity we only use a plain distinction in each quality, which results in a matrix breakdown of service levels as shown in Figure~\ref{fig:qosdim}. More sophisticated differentiations apply analogously.

Service classes are assigned to flows according to a list of prefixes that are marked with a traffic class and maintained by each node.
Incoming Interest and Data messages are then mapped onto traffic classes by applying a longest prefix match against this list of traffic identifiers. Traffic class names may look as follows:
\vspace{0.25cm}
\begin{center}
\noindent\begin{tabular}{@{}*{2}{p{.5\columnwidth}@{}}} \toprule
  \multicolumn{1}{c}{Traffic Class} & \multicolumn{1}{c}{Priority} \\ \midrule
  /HK/ACM/ICN & \multicolumn{1}{c}{$<$Reliable, Regular$>$}  \\
  /HK/ACM/ICN/site/A/alarm & \multicolumn{1}{c}{$<$Reliable, Prompt$>$} \\
  /HK/ACM/ICN/site/B/temp & \multicolumn{1}{c}{$<$Regular, Prompt$>$} \\
  \bottomrule
\end{tabular}
\end{center}
\vspace{0.25cm}
\noindent Given this example, Interest and Data messages for the name prefix \textit{/HK/ACM/ICN/site/C} would map to the class of \textit{/HK/ACM/ICN}.

In this work, we assume such lists deployed at all nodes within a network domain. The distribution and maintenance of QoS configurations may be use case dependent and introduce a complex dynamic. Its generic treatment is beyond the scope of this work, which focuses on  a lightweight flow-to-priority mapping for the purpose of quantifying the benefits of QoS resource management.  

It may be noted that fine-grained service differentiation within a complex name hierarchy can result in large QoS prefix tables that consume significant memory. For the constrained IoT use case, though, we argue that sensor readings and actuator settings are machine type communications (MTC) with (short) names configurable according to processing needs. Hence QoS overheads should easily comply to resource constraints.

\subsection{Manageable Resources}%
\label{subsec:qosprios}

\subsubsection{Link Layer}
 The link layer manages access to the media and provides space to buffer packets. In low-power wireless networks, 
media access times are highly susceptible to media saturation and buffer spaces are small.
While time slotted technologies such as the  IEEE~802.15.4e TSCH mode and Bluetooth Low Energy access 
media in a deterministically scheduled manner, the unslotted CSMA/CA version of IEEE~802.15.4 or long-range radios such as LoRa are more susceptible to packet collisions between neighboring nodes.

In the IoT, content producers that generate sensor readings may produce egress traffic at a rate that is much
higher than the average media access time.
In addition, nodes may further need to forward ingress traffic in multi-hop scenarios.
For this purpose, buffering egress traffic is necessary to cope with traffic spikes.

Queuing and buffering take the same role in ICN as in the IP world. Class-based forwarding queues will process packets of the \emph{prompt} flow class before packets with a regular priority while buffer space will prevent packet drops. It should be noted, though, that rapidly forwarded packets in NDN will also quickly satisfy PIT entries and thereby free forwarding resources for unprioritized traffic on the same path.

\subsubsection{Pending Interest Table}
The Pending Interest Table (PIT) enables the stateful forwarding plane of CCNx and NDN and thereby governs the flows in the network.
The size of the PIT resource effectively dictates the maximum number of simultaneous open requests and the coherence of PIT entries along a path determines whether flows can propagate without barriers.
In normal NDN operation, PIT state is allocated when an Interest message is processed, and
it is removed in two scenarios. Either a returning Data message consumes the PIT state,
or a timeout after a succession of retransmissions clears the state.

 PIT saturation is common even in overprovisioned networks, but is far more likely to occur
in IoT deployments. Limited RAM resources, slow processing power, delayed media access, and packet loss
in low-power networks with intermittent connectivity can all cause the PIT to reach its maximum capacity.

The typical way of handling incoming Interest messages at a saturated PIT is to drop them in order to avoid
cancelling active but incomplete request operations. The penalty for dropping such Interests is an increase
in latency due to retransmissions, which usually happen on the scale of seconds.
To avoid high latencies for time-sensitive traffic flows, an elaborate PIT eviction strategy is necessary, which 
accounts for \one unhindered forwarding of prioritized Data, and harmonizes with \two the retransmission mechanisms of the regular NDN.

\subsubsection{Content Store}
Due to memory constraints, the Content Store (CS) is typically small, 
necessitating heuristics for deciding what content to
cache at which node instead of indiscriminately caching all incoming
content. There is a wealth of existing research on how to make the most
efficient use of limited CS space, with a number of different strategies
employing various heuristics to decide whether or not to cache incoming
content (see Section~\ref{subsec:relwork_caching}). This aspect of
caching is called the \emph{caching decision strategy}.

The introduction of traffic flow priorities adds an additional
dimension to the caching decision. Regardless of which specific caching
decision strategy is employed, content marked as \emph{reliable} should
always be cached as it is imperative that this content is available
throughout the network.
Thus, reception of \emph{reliable} content should not trigger the
caching decision strategy; instead, control should be handed directly to
the cache replacement strategy (see below). The question whether
\emph{prompt} content should be cached with a higher priority than
content with \emph{regular} latency requirements is not as clear-cut.
Caching \emph{prompt} content with higher priority would have a positive
effect on future transmissions of that content object (either by
retransmission of the original request or by new requests) and thus have
a positive effect on latency, although the potential gain in this aspect
is dependent on path length. Any content that is marked as
\emph{regular} in both QoS dimensions should be treated as normal; in
other words, the caching decision strategy is consulted.

After a node has decided to cache a new content object, an additional
step may have to be taken in case the CS is at capacity. This aspect of
caching is the \emph{cache replacement strategy}. In most cases, CS
contents will be replaced using a simple heuristic such as Least
Recently Used (LRU). However, once again the introduction of traffic
flow priorities adds an additional dimension to this decision.

In general, incoming content should not replace content of a higher
priority. Therefore, content with \emph{regular} latency requirements
should not replace \emph{prompt} content and content with
\emph{regular} reliability should not replace \emph{reliable} content.
When it comes to the correlation between latency and reliability, the
primary goal of the CS should be to ensure content availability, which
places a stronger emphasis on the reliability aspect. Thus,
\emph{reliable} content with \emph{regular} latency should be able to
replace \emph{prompt} content if no other content is eligible to be
replaced. If all content is of the same priority class, regular
replacement rules (e.g., LRU) should apply.

In probabilistic caching, as introduced in
Section~\ref{subsec:relwork_caching}, each node caches incoming content
according to a certain probability $p$. Regardless of how exactly $p$ is
determined (whether statically or by one of the dynamic methods
discussed in Section~\ref{subsec:relwork_caching}), the probabilistic
approach may be refined by differentiating between two separate
probabilities $p_{reg}$ for regular content and $p_{rel}$ for reliable
content, with $p_{rel} > p_{reg}$. This has the effect that a CS at each node
 will have different contents, thus contributing to CS diversity
across the network by making a larger range of content available as
cached copies, while giving consideration to service classes ensures
that higher-priority content is still treated preferentially.

\subsection{Distributed QoS Management}

We are now ready to present our approaches to distributed resource management for supporting QoS in ICN. 
The corresponding mechanisms  fall into three
classes, depending on the level of interdependence  between
resources on the same device or between devices:

\subsubsection{Locally Isolated Decisions}

The straightforward allocation of independent resources to packet forwarding follows three simple rules: 

\begin{description}
	\item[Prioritized forwarding] applies to flows marked as \emph{prompt}.

	\item[Cache] (re-)placement decisions obey the priority order  \emph{reliable} (highest) to  \emph{regular} (lowest). In the presence of probabilistic caching strategies, the weights are set accordingly. 

	\item[PIT management] operates in favor of rapid packet forwarding, so PIs enter the PIT in the order \emph{prompt} (highest)  to  \emph{regular} (lowest). Newly arriving Interests that meet a PIT saturated with entries of equal or higher priority will be dropped.
\end{description}

\subsubsection{Local Resource Correlations}

These are decisions that entail interaction between mechanisms on the
same device (intra-device correlations). This includes the correlation
between the caching decision and cache replacement strategies, where
e.g.\ the caching decision may pre-empt the cache replacement decision
in the case of \emph{reliable} flows, while the cache replacement
decision may drop content even in the case of a positive caching
decision if no content of the same or lower service class can be
replaced. In detail we take the following steps:
\begin{description}
	\item[w/ PIT entry] If arriving Data meets a valid PIT entry, Data is forwarded according to priorities \emph{and cached} with priority, if marked as \emph{reliable}. In the case  of exhausted prioritized forwarding queue, \emph{prompt} traffic will be cached with the highest priority.  

	\item[w/o PIT entry] If arriving Data meets no valid PIT entry, cache placement will still be initiated for \emph{prompt} and  \emph{reliable} data in subsequent order. 
          For probabilistic caching, weights are adjusted accordingly.
\end{description}

In balanced, unconstrained NDN networks, returning regular Data meets open PIT states.
For saturated PITs, however, PIT entries may time out quickly, or resource management may enforce eviction of PIT entries in favor of other requests.
Allowing Data without corresponding PIT entries to be cached may introduce the threat of cache poisoning attacks.
However, a simple rate limiting on incoming Data packets and a reduced cache time for these CS entries may reduce the attack surface in our constrained environment.
Further analysis of related effects is left to future work.

\subsubsection{Distributed Resource Coordination}

Such mechanisms affect resources across multiple or all devices
in the network (inter-device correlations). These include maintaining
PIT coherence by ensuring that all nodes apply uniform QoS mechanisms
when replacing content of different service classes, as well as
achieving CS diversity by introducing probabilistic caching based on
priority classes. In our system, distributed coordination is achieved as follows.

\begin{description}
	\item[PIT coherence] is increased by  applying the same PIT eviction strategy at all nodes, i.e., evict \emph{regular} before  \emph{reliable} before  \emph{prompt}.
	\item[Cache efficiency] increases with probabilistic caching using the coordination of equal cache weights. It is noteworthy that probabilistic caching reduces the risk of starvation for low priority content due to higher cache diversity, even if high priority flows dominate the network.
\end{description}


\section{Experimental Evaluations}\label{sec:eval}
\subsection{Implementation and Experiment Setup}
\subsubsection{Software and Hardware Platforms}
We conduct our experiments on the FIT IoT-Lab testbed using typical class~2~\cite{RFC-7228} IoT devices that feature an ARM Cortex-M3 MCU with 64~kB RAM and 512~kB ROM.\@
Each device further contains an Atmel AT86RF231~\cite{a-lptzi-09} 2.4~GHz transceiver to operate on the IEEE~802.15.4 radio.

All devices run on RIOT~OS~\cite{bhgws-rotoi-13} version \texttt{2019.04} with the integrated NDN network stack CCN-lite~\cite{ccn-lite},  which we extended with our QoS management scheme. In addition to the PIT and CS management strategies, a very lightweight prioritized forwarding was implemented using a single packet double-buffer that allows for pairwise packet re-ordering. We also note that network stack performance usually exceeds link speed.
While IEEE~802.15.4 provides a theoretical maximum of 250~kbit/s, I/O and data processing in the network stack is at least one order of magnitude faster~\cite{lkhpg-cwemr-18}.

Following the large-scale deployment in~\cite{gklp-ncmcm-18}, we configure a maximum number of four retransmissions and
a retransmission interval of two seconds for CCN-lite.
To analyze our approach under different levels of network saturation, we configure the maximum capacities of PIT and CS to
range between 5 and 30 elements.
Notably, the estimated RAM usage for 30 PIT entries and 30 CS elements with name lengths of $\approx$ 32~bytes is already approximately 11~KiB, which is around 17\% of the total available RAM for our hardware platform.

\subsubsection{Topology Setup}
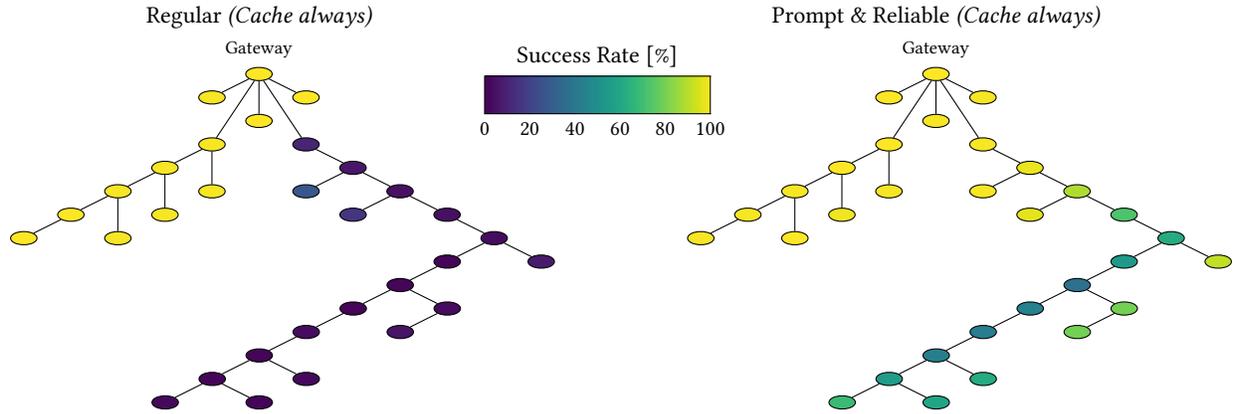
\begin{figure*}[t]
  \centering
\begin{tikzpicture}[>={Latex},/pgfplots/colormap/viridis]
  \tikzset{
    router/.style={fill=Pastel1-A,draw=black,circle,inner sep=2.5pt},
    gw/.style={router},
    success/.style={
            /pgfplots/color of colormap={#1*10},
            draw=black,
            fill=.,
        },
  }

  \pgfplotscolorbardrawstandalone[colorbar horizontal,
  point meta min=0,point meta max=100,
  colorbar style={
    at={(3cm,1.0cm)},
    width=3cm,
    title={Success Rate [\%]},title style={yshift=-0.2cm},
    x tick label style={major tick length=0pt,font=\small,align=center},
  },]
  \begin{scope}[xscale=1.25,yscale=1.25,
    yshift=-83,every node/.append style={
      yslant=0.5,xslant=-1},yslant=0.5,xslant=-1
    ]
    \node[gw,success=100] (m3a-96) at (3.5,3.5) {};
    \node[router,success=99] (m3a-99) at ([xshift=-0.5cm,yshift=0.0cm]m3a-96) {};
    \node[router,success=100] (m3a-104) at ([xshift=0.0cm,yshift=-0.5cm]m3a-96) {};
    \node[router,success=100] (m3a-112) at ([xshift=-0.5cm,yshift=-0.5cm]m3a-96) {};
    \node[router,success=99] (m3a-115) at ([xshift=-1.0cm,yshift=-0.5cm]m3a-96) {};
    \node[router,success=10] (m3a-94) at ([xshift=-0.5cm,yshift=-1cm]m3a-96) {};
    \node[router,success=99] (m3a-129) at ([xshift=-0.5cm,yshift=0cm]m3a-115) {};
    \node[router,success=99] (m3a-149) at ([xshift=-0.5cm,yshift=0cm]m3a-129) {};
    \node[router,success=100] (m3a-166) at ([xshift=-0.5cm,yshift=0cm]m3a-149) {};
    \node[router,success=100] (m3a-175) at ([xshift=-0.5cm,yshift=0cm]m3a-166) {};
    \node[router,success=99] (m3a-122) at ([xshift=-0.5cm,yshift=-0.5cm]m3a-115) {};
    \node[router,success=99] (m3a-142) at ([xshift=-0.5cm,yshift=-0.5cm]m3a-129) {};
    \node[router,success=99] (m3a-157) at ([xshift=-0.5cm,yshift=-0.5cm]m3a-149) {};
    \node[router,success=6] (m3a-90) at ([xshift=0cm,yshift=-0.5cm]m3a-94) {};
    \node[router,success=27] (m3a-86) at ([xshift=-0.5cm,yshift=0cm]m3a-90) {};
    \node[router,success=4] (m3a-80) at ([xshift=0cm,yshift=-0.5cm]m3a-90) {};
    \node[router,success=4] (m3a-70) at ([xshift=0cm,yshift=-0.5cm]m3a-80) {};
    \node[router,success=15] (m3a-76) at ([xshift=-0.5cm,yshift=0cm]m3a-80) {};
    \node[router,success=3] (m3a-1) at ([xshift=0cm,yshift=-0.5cm]m3a-70) {};
    \node[router,success=7] (m3a-6) at ([xshift=0cm,yshift=-0.5cm]m3a-1) {};
    \node[router,success=1] (m3a-9) at ([xshift=-0.5cm,yshift=0cm]m3a-1) {};
    \node[router,success=1] (m3a-15) at ([xshift=-0.5cm,yshift=0cm]m3a-9) {};
    \node[router,success=1] (m3a-20) at ([xshift=-0.5cm,yshift=0cm]m3a-15) {};
    \node[router,success=3] (m3a-30) at ([xshift=-0.5cm,yshift=0cm]m3a-20) {};
    \node[router,success=3] (m3a-24) at ([xshift=0cm,yshift=-0.5cm]m3a-15) {};
    \node[router,success=5] (m3a-34) at ([xshift=-0.5cm,yshift=0cm]m3a-24) {};
    \node[router,success=1] (m3a-40) at ([xshift=-0.5cm,yshift=0cm]m3a-30) {};
    \node[router,success=2] (m3a-47) at ([xshift=0cm,yshift=-0.5cm]m3a-40) {};
    \node[router,success=1] (m3a-54) at ([xshift=-0.5cm,yshift=0cm]m3a-40) {};
    \node[router,success=1] (m3a-60) at ([xshift=0cm,yshift=-0.5cm]m3a-54) {};
    \node[router,success=2] (m3a-65) at ([xshift=-0.5cm,yshift=0cm]m3a-54) {};
    \path[]
    (m3a-96) edge (m3a-99)
    (m3a-96) edge (m3a-104)
    (m3a-96) edge (m3a-112)
    (m3a-96) edge (m3a-115)
    (m3a-96) edge (m3a-94)
    (m3a-115) edge (m3a-129)
    (m3a-129) edge (m3a-149)
    (m3a-149) edge (m3a-166)
    (m3a-166) edge (m3a-175)
    (m3a-115) edge (m3a-122)
    (m3a-129) edge (m3a-142)
    (m3a-149) edge (m3a-157)
    (m3a-94) edge (m3a-90)
    (m3a-90) edge (m3a-80)
    (m3a-86) edge (m3a-90)
    (m3a-80) edge (m3a-70)
    (m3a-80) edge (m3a-76)
    (m3a-70) edge (m3a-1)
    (m3a-1) edge (m3a-6)
    (m3a-1) edge (m3a-9)
    (m3a-15) edge (m3a-9)
    (m3a-20) edge (m3a-15)
    (m3a-30) edge (m3a-20)
    (m3a-15) edge (m3a-24)
    (m3a-24) edge (m3a-34)
    (m3a-30) edge (m3a-40)
    (m3a-40) edge (m3a-54)
    (m3a-40) edge (m3a-47)
    (m3a-54) edge (m3a-60)
    (m3a-54) edge (m3a-65)
    ;
  \end{scope}
  \node[align=center,anchor=south,font=\footnotesize] at ([yshift=0.1cm]m3a-96) {Gateway};
  \node[align=center,anchor=south] at ([yshift=0.5cm]m3a-96) {Regular \it (Cache always)};
  \begin{scope}[xshift=9cm,xscale=1.25,yscale=1.25,
    yshift=-83,every node/.append style={
      yslant=0.5,xslant=-1},yslant=0.5,xslant=-1,
    ]
    \node[gw,success=100] (m3a-96) at (3.5,3.5) {};
    \node[router,success=100] (m3a-99) at ([xshift=-0.5cm,yshift=0.0cm]m3a-96) {};
    \node[router,success=99] (m3a-104) at ([xshift=0.0cm,yshift=-0.5cm]m3a-96) {};
    \node[router,success=99] (m3a-112) at ([xshift=-0.5cm,yshift=-0.5cm]m3a-96) {};
    \node[router,success=100] (m3a-115) at ([xshift=-1.0cm,yshift=-0.5cm]m3a-96) {};
    \node[router,success=99] (m3a-94) at ([xshift=-0.5cm,yshift=-1cm]m3a-96) {};
    \node[router,success=100] (m3a-129) at ([xshift=-0.5cm,yshift=0cm]m3a-115) {};
    \node[router,success=99] (m3a-149) at ([xshift=-0.5cm,yshift=0cm]m3a-129) {};
    \node[router,success=99] (m3a-166) at ([xshift=-0.5cm,yshift=0cm]m3a-149) {};
    \node[router,success=99] (m3a-175) at ([xshift=-0.5cm,yshift=0cm]m3a-166) {};
    \node[router,success=99] (m3a-122) at ([xshift=-0.5cm,yshift=-0.5cm]m3a-115) {};
    \node[router,success=98] (m3a-142) at ([xshift=-0.5cm,yshift=-0.5cm]m3a-129) {};
    \node[router,success=100] (m3a-157) at ([xshift=-0.5cm,yshift=-0.5cm]m3a-149) {};
    \node[router,success=97] (m3a-90) at ([xshift=0cm,yshift=-0.5cm]m3a-94) {};
    \node[router,success=99] (m3a-86) at ([xshift=-0.5cm,yshift=0cm]m3a-90) {};
    \node[router,success=88] (m3a-80) at ([xshift=0cm,yshift=-0.5cm]m3a-90) {};
    \node[router,success=72] (m3a-70) at ([xshift=0cm,yshift=-0.5cm]m3a-80) {};
    \node[router,success=96] (m3a-76) at ([xshift=-0.5cm,yshift=0cm]m3a-80) {};
    \node[router,success=61] (m3a-1) at ([xshift=0cm,yshift=-0.5cm]m3a-70) {};
    \node[router,success=91] (m3a-6) at ([xshift=0cm,yshift=-0.5cm]m3a-1) {};
    \node[router,success=54] (m3a-9) at ([xshift=-0.5cm,yshift=0cm]m3a-1) {};
    \node[router,success=37] (m3a-15) at ([xshift=-0.5cm,yshift=0cm]m3a-9) {};
    \node[router,success=44] (m3a-20) at ([xshift=-0.5cm,yshift=0cm]m3a-15) {};
    \node[router,success=42] (m3a-30) at ([xshift=-0.5cm,yshift=0cm]m3a-20) {};
    \node[router,success=79] (m3a-24) at ([xshift=0cm,yshift=-0.5cm]m3a-15) {};
    \node[router,success=79] (m3a-34) at ([xshift=-0.5cm,yshift=0cm]m3a-24) {};
    \node[router,success=43] (m3a-40) at ([xshift=-0.5cm,yshift=0cm]m3a-30) {};
    \node[router,success=61] (m3a-47) at ([xshift=0cm,yshift=-0.5cm]m3a-40) {};
    \node[router,success=55] (m3a-54) at ([xshift=-0.5cm,yshift=0cm]m3a-40) {};
    \node[router,success=59] (m3a-60) at ([xshift=0cm,yshift=-0.5cm]m3a-54) {};
    \node[router,success=68] (m3a-65) at ([xshift=-0.5cm,yshift=0cm]m3a-54) {};
    \path[]
    (m3a-96) edge (m3a-99)
    (m3a-96) edge (m3a-104)
    (m3a-96) edge (m3a-112)
    (m3a-96) edge (m3a-115)
    (m3a-96) edge (m3a-94)
    (m3a-115) edge (m3a-129)
    (m3a-129) edge (m3a-149)
    (m3a-149) edge (m3a-166)
    (m3a-166) edge (m3a-175)
    (m3a-115) edge (m3a-122)
    (m3a-129) edge (m3a-142)
    (m3a-149) edge (m3a-157)
    (m3a-94) edge (m3a-90)
    (m3a-90) edge (m3a-80)
    (m3a-86) edge (m3a-90)
    (m3a-80) edge (m3a-70)
    (m3a-80) edge (m3a-76)
    (m3a-70) edge (m3a-1)
    (m3a-1) edge (m3a-6)
    (m3a-1) edge (m3a-9)
    (m3a-15) edge (m3a-9)
    (m3a-20) edge (m3a-15)
    (m3a-30) edge (m3a-20)
    (m3a-15) edge (m3a-24)
    (m3a-24) edge (m3a-34)
    (m3a-30) edge (m3a-40)
    (m3a-40) edge (m3a-54)
    (m3a-40) edge (m3a-47)
    (m3a-54) edge (m3a-60)
    (m3a-54) edge (m3a-65)
    ;
  \end{scope}
  \node[align=center,anchor=south,font=\footnotesize] at ([yshift=0.1cm]m3a-96) {Gateway};
  \node[align=center,anchor=south] at ([yshift=0.5cm]m3a-96) {Prompt \& Reliable \it (Cache always)};
\end{tikzpicture}

%
  \caption{Nodal success rates for \textit{Scenario 1} using regular traffic (left) and \textit{reliable} actuator traffic (right).}%
  \label{fig:topology}
\end{figure*}
The testbed provides access to multiple sites with varying numbers of M3 IoT devices and network characteristics.
We deploy our applications on the \textit{Grenoble} site, as this supports significantly complex multi-hop paths.
We choose 31 M3 devices arbitrarily, where one device acts as a gateway node and the other 30 devices act as sensors and actuators.
Since convergecast is the most predominant traffic pattern in common IoT scenarios, we arrange our devices to form a
\textit{Destination Oriented Directed Acyclic Graph (DODAG)} that is rooted at the gateway node.
Approximately 60\% of the nodes are reachable from the root within 4--5 hops, while the remaining devices have path lengths up to 12 hops. The topology is visualized in Figure~\ref{fig:topology}.
Extended left and right wings in the routing topology result from long hallways in the \textit{Grenoble} site.

\subsubsection{Scenarios}
We want to quantify the efficiency of our QoS management scheme in saturated multi-hop deployments that display typical traffic patterns.
With this in view, we analyze our approach under two different scenarios.

\paragraph{Scenario 1: Mixed Sensors and Actuators}
The gateway node requests temperature readings from the 30 sensor nodes every 10~s with $\pm$ 2~s jitter interval.
Thus, on average, the request rate at the gateway approximates to 3~\textit{packets/s}  and
including the reception rate of responses, the gateway handles 6~\textit{packets/s}.
The naming scheme for each request consists of a prefix, a device-specific $node\ id$, and an increasing sequence number.
We refer to this traffic equally as \textit{sensor readings} or \textit{gateway initiated traffic}.

In addition, all 30 devices further act as actuators that periodically request a device-specific state from the gateway node every 5~s with $\pm$ 1~s jitter.
This yields a request reception rate of 6~\textit{packets/s} on the root node.
The naming scheme for these requests similarly consists of a prefix, a device-specific $node\ id$, and an increasing sequence number.
We name this traffic \textit{actuator initiated traffic}.

\paragraph{Scenario 2: Sensing and Lighting Control}
In the second scenario, we change the role of actuators but leave the sensor readings unchanged. 
Instead of independent actuators that receive disjoint instructions, we envision a scenario of lighting in which groups of fixtures switch lights in a coordinated way. Hence, these groups receive identical commands and caching becomes applicable.   

To explore the event space by mixing group memberships, our experiments proceed as follows.
For each request, an actuator randomly joins one out of five possible groups.
The naming scheme for such requests is changed to include the selected $group\ id$, instead of a device-specific $node\ id$.
Besides naming, we use the same request parameters for the \textit{actuator initiated traffic} as in the first scenario.
We repeat this process  240 times for each configuration in order to explore the state space of unevenly distributed groups and converge statistics.

\subsubsection{Caching Parameters}
We consider the caching procedure to consist of two fundamental steps: \one caching decision, and \two cache replacement.
In our experiments, we use two different caching decision strategies and one cache replacement strategy.

The first decision strategy is to \textit{cache always} incoming Data packets, with the restriction for regular traffic
that Data packets without a corresponding PIT entry are dropped and not cached.
For \emph{reliable} traffic, we adjust this strategy, such that Data packets without PIT entries are still considered
for caching.
The second decision strategy is to \textit{cache probabilistically}.
Every node caches incoming Data packets with a probability $p_{reg}$ of 30\% for regular traffic,
and with a probability $p_{rel}$ of 70\% for \emph{reliable} traffic.
For a saturated CS, our cache replacement strategy evicts content store elements using the \textit{least recently used (LRU)} policy.

\subsection{Results}
\begin{figure*}[t]
  \centering
  \tikzexternalenable
\tikzsetnextfilename{all-successes}
\begin{tikzpicture}
  \begin{groupplot}[
    group style={
      group size=6 by 2,
      horizontal sep=0.20cm,
      vertical sep=1.25cm,
    },
    xmin=3.0,xmax=32.0,ymin=0.5,ymax=12.5,
    width=0.22\textwidth,
    height=0.22\textwidth,
    colormap/viridis,
    enlargelimits=false,
    axis on top,
    point meta min=0,
    point meta max=100,
    x tick label style={major tick length=0pt},
    xtick = {5,10,15,20,25,30},
    ytick = {2,4,6,8,10,12},
    xticklabels={5, 10, 15, 20, 25, 30},
    xticklabel style = {align=center},
    title style={anchor=north,yshift=0.75cm,},
    ]

    \nextgroupplot[align=center,title={Regular\\\it (Cache always)}]
    \addplot [matrix plot*,point meta=explicit] file [meta=index 2] {data/sa_config1_CS5-10s5s-asuccesses.csv};
    \nextgroupplot[yticklabels={},align=center,title={Prompt \& Rel.\\\it (Cache always)}]
    \addplot [matrix plot*,point meta=explicit] file [meta=index 2] {data/sa_config3_CS5-10s5s-asuccesses.csv};
    \nextgroupplot[yticklabels={},align=center,title={Prompt \& Rel.\\\it (Cache probab.)}]
    \addplot [matrix plot*,point meta=explicit] file [meta=index 2] {data/sa_config15_CS5-10s5s-asuccesses.csv};
    \nextgroupplot[yticklabels={},align=center,title={Regular\\\it (Cache always)}]
    \addplot [matrix plot*,point meta=explicit] file [meta=index 2] {data/sa_config1_CS5-10s5s-ssuccesses.csv};
    \nextgroupplot[yticklabels={},align=center,title={Prompt \& Rel.\\\it (Cache always)}]
    \addplot [matrix plot*,point meta=explicit] file [meta=index 2] {data/sa_config3_CS5-10s5s-ssuccesses.csv};
    \nextgroupplot[yticklabels={},align=center,title={Prompt \& Rel.\\\it (Cache probab.)},colorbar,colorbar style={anchor=north west,xshift=-0.3cm,ylabel={Success Rate [\%]},ylabel style={yshift=0.15cm},height=2.54*\pgfkeysvalueof{/pgfplots/parent axis width}}]
    \addplot [matrix plot*,point meta=explicit] file [meta=index 2] {data/sa_config15_CS5-10s5s-ssuccesses.csv};
    \nextgroupplot[align=center]
    \addplot [matrix plot*,point meta=explicit] file [meta=index 2] {data/sa_config7_PIT5-10s5s-asuccesses.csv};
    \nextgroupplot[yticklabels={},align=center]
    \addplot [matrix plot*,point meta=explicit] file [meta=index 2] {data/sa_config8_PIT5-10s5s-asuccesses.csv};
    \nextgroupplot[yticklabels={},align=center]
    \addplot [matrix plot*,point meta=explicit] file [meta=index 2] {data/sa_config9_PIT5-10s5s-asuccesses.csv};
    \nextgroupplot[yticklabels={},align=center]
    \addplot [matrix plot*,point meta=explicit] file [meta=index 2] {data/sa_config7_PIT5-10s5s-ssuccesses.csv};
    \nextgroupplot[yticklabels={},align=center]
    \addplot [matrix plot*,point meta=explicit] file [meta=index 2] {data/sa_config8_PIT5-10s5s-ssuccesses.csv};
    \nextgroupplot[yticklabels={},align=center,]
    \addplot [matrix plot*,point meta=explicit] file [meta=index 2] {data/sa_config9_PIT5-10s5s-ssuccesses.csv};

  \end{groupplot}
  \path (group c1r1.north west) -- (group c3r1.north east) node[midway,above=1.0cm,anchor=south] {Actuator Initiated Traffic};
  \path (group c4r1.north west) -- (group c6r1.north east) node[midway,above=1.0cm,anchor=south] {Gateway Initiated Traffic};
  \path (group c1r1.south west) -- (group c6r1.south east) node[midway,below=0.5cm,anchor=north] {PIT Size [Maximum \# of Entries]};
  \path (group c1r2.south west) -- (group c6r2.south east) node[midway,below=0.5cm,anchor=north] {CS Size [Maximum \# of Entries]};
  \path (group c1r2.south west) -- (group c1r1.north west) node[midway,sloped,above=0.5cm,anchor=south] {Rank [Hops]};
  \node[rotate=90,anchor=south east] at ([xshift=-0.57cm]group c1r1.north west) {\it Scenario 1};
  \node[rotate=90,anchor=south west] at ([xshift=-0.57cm]group c1r2.south west) {\it Scenario 2};
\end{tikzpicture}
\tikzexternaldisable

%
  \caption{Success rates per rank for \textit{Scenario 1} and \textit{Scenario 2} using varying PIT and CS sizes.}%
  \label{fig:scenario1and2_successes}
\end{figure*}
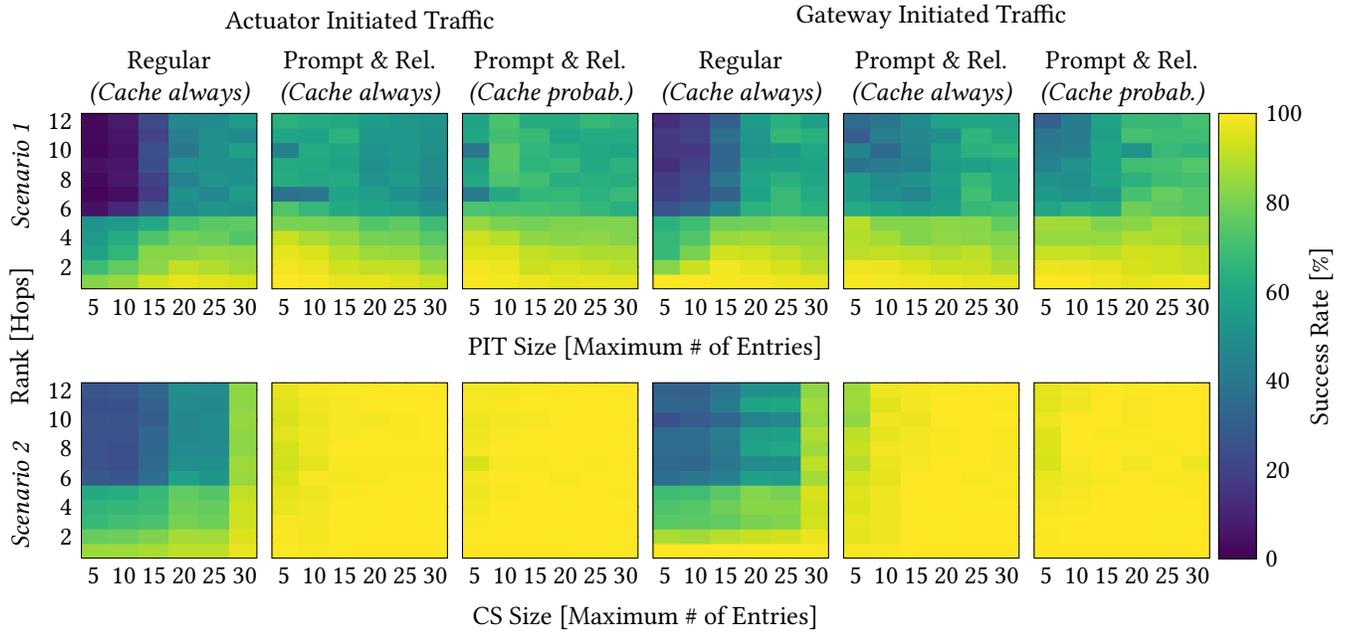

\subsubsection{Analyses and metrics}
We analyze the network utilization for gateway and actuator initiated traffic separately in the presented scenarios with and without the proposed QoS features.
Furthermore, we measure the \textit{success rates}, \textit{goodputs} and \textit{time to completion} for each node in the topology.

\subsubsection{Scenario 1: Mixed Sensors and Actuators}
In this scenario, sensor and actuator data are device-specific and thus only single destinations benefit from on-path caching during the narrow window of Interest retransmissions.

\paragraph{Success rates}
In our first experiment we focus on the nodal success rates using the first scenario with a PIT and CS limitation of 5 entries.
The gateway is configured with a PIT limitation of 50.
Figure~\ref{fig:topology} shows the resulting success rates for \one the regular operation of NDN on the left hand side,
and \two a setup with prioritized actuator initiated traffic using the \textit{prompt} and \textit{reliable} QoS service levels on the right hand side.
The success rates per node are color coded and range from 0\% (purple) to 100\% (yellow).

Figure~\ref{fig:topology} clearly depicts huge differences in success rates for both configurations.
In the normal NDN operation, nodes close to the gateway, as well as the left wing of the topology, achieve 100\% success rates.
Strikingly, the right wing exhibits major network stress, with nearly all actuators having a success rate below 10\% due to  PIT overflows.
Conversely, with QoS service levels enabled, the right wing shows much enhanced network performance, which results in overall higher success.
With this configuration, 70--100\% of the packets arrive at actuators close to the gateway, and 40--70\% at more distant nodes.
Leaf nodes farther away show a greater improvement in success rates than forwarding ancestor nodes.

This striking example nicely illustrates the positive effect of PIT correlation under QoS. While in the regular case Interests  at nodes with exhausted PIT are discarded as they randomly arrive, the QoS marking preselects those requests that are prioritized  throughout the network, leading to fewer retransmissions and a more efficient use of the overall PIT space available in the network.

We now dig deeper into network reliability and examine the success rates for a range of  PIT sizes plotted against the node ranks (first row of Figure~\ref{fig:scenario1and2_successes}).
As previously observed, the success rate for \textit{regular} actuator traffic collapses at higher ranks.
The gateway traffic performs similarly poor.
Increasing the maximum PIT size gradually improves the nodal success rates.
In addition to the \textit{regular} traffic, we further show actuator traffic with \textit{prompt \& reliable} QoS service levels.
Overall, the setups with prioritized traffic show great improvements for the smaller PIT sizes,
while the \textit{probabilistic} caching decision strategy performs slightly better than the \textit{always} strategy.
A surprising effect is observed with a PIT size of 30: the success rates for all configurations decline slightly for lower ranks.
This is caused by an increased retransmission overhead per node, resulting in link saturation.

\begin{figure*}[t]
  \centering
\begin{tikzpicture}
  \begin{groupplot}[
    group style={
      group size=2 by 2,
      horizontal sep=0.25cm,
      vertical sep=0.25cm,
    },
    height=3.5cm,
    width=0.53\textwidth,
    no markers,
    xtick align=inside,
    xtick pos=left,
    xtick={5,10,15,20},
    axis on top,
    title style={yshift=-0.2cm},
    ]
    \nextgroupplot[xticklabels={},ymin=0,ymax=650,xmin=0,xmax=20,title={Outgoing Requests at Gateway}]
    \addplot[draw=black,fill=Pastel1-B] table [x index=0, y index=1] {data/baseline_sa_config1_pit5_CS5.csv} \closedcycle;
    \draw[draw=black,densely dashed,thick] ({axis cs:9,0}|-{rel axis cs:0,1}) -- ({axis cs:9,0}|-{rel axis cs:0,0});
    \draw[draw=black,densely dashed,thick] ({axis cs:19,0}|-{rel axis cs:0,1}) -- ({axis cs:19,0}|-{rel axis cs:0,0});
    \node[font=\footnotesize,align=center] at ({axis cs:4.65,0}|-{rel axis cs:0,0.38}) {Without Actuator Traffic};
    \node[font=\footnotesize,align=center] at ({axis cs:14,0}|-{rel axis cs:0,0.38}) {With Actuator Traffic};
    \node[font=\footnotesize,align=center,anchor=north west] at ({rel axis cs:0.01,1}) {Regular\\\it Always / LRU};
    \nextgroupplot[yticklabels={},xticklabels={},ymin=0,ymax=650,xmin=0,xmax=20,title={Incoming Responses at Gateway}]
    \addplot [black,fill=Pastel1-B] table [x index=0, y index=2] {data/baseline_sa_config1_pit5_CS5.csv} \closedcycle;
    \draw[draw=black,densely dashed,thick] ({axis cs:9,0}|-{rel axis cs:0,1}) -- ({axis cs:9,0}|-{rel axis cs:0,0});
    \draw[draw=black,densely dashed,thick] ({axis cs:19,0}|-{rel axis cs:0,1}) -- ({axis cs:19,0}|-{rel axis cs:0,0});
    \node[font=\footnotesize,align=center] at ({axis cs:4.65,0}|-{rel axis cs:0,0.38}) {Without Actuator Traffic};
    \node[font=\footnotesize,align=center] at ({axis cs:14,0}|-{rel axis cs:0,0.38}) {With Actuator Traffic};
    \node[font=\footnotesize,align=center,anchor=north west] at ({rel axis cs:0.01,1}) {Regular\\\it Always / LRU};

    \nextgroupplot[ymin=0,ymax=650,xmin=0,xmax=20]
    \addplot [black,fill=Pastel1-B] table [x index=0, y index=1] {data/baseline_sa_config3_pit5_CS5.csv} \closedcycle;
    \draw[draw=black,densely dashed,thick] ({axis cs:9,0}|-{rel axis cs:0,1}) -- ({axis cs:9,0}|-{rel axis cs:0,0});
    \draw[draw=black,densely dashed,thick] ({axis cs:19,0}|-{rel axis cs:0,1}) -- ({axis cs:19,0}|-{rel axis cs:0,0});
    \node[font=\footnotesize,align=center] at ({axis cs:4.65,0}|-{rel axis cs:0,0.38}) {Without Actuator Traffic};
    \node[font=\footnotesize,align=center] at ({axis cs:14,0}|-{rel axis cs:0,0.38}) {With Actuator Traffic};
    \node[font=\footnotesize,align=center,anchor=north west] at ({rel axis cs:0.01,1}) {Prompt \& Reliable\\\it Always / LRU};

    \nextgroupplot[yticklabels={},ymin=0,ymax=650,xmin=0,xmax=20]
    \addplot [black,fill=Pastel1-B] table [x index=0, y index=2] {data/baseline_sa_config3_pit5_CS5.csv} \closedcycle;
    \draw[draw=black,densely dashed,thick] ({axis cs:9,0}|-{rel axis cs:0,1}) -- ({axis cs:9,0}|-{rel axis cs:0,0});
    \draw[draw=black,densely dashed,thick] ({axis cs:19,0}|-{rel axis cs:0,1}) -- ({axis cs:19,0}|-{rel axis cs:0,0});
    \node[font=\footnotesize,align=center] at ({axis cs:4.65,0}|-{rel axis cs:0,0.38}) {Without Actuator Traffic};
    \node[font=\footnotesize,align=center] at ({axis cs:14,0}|-{rel axis cs:0,0.38}) {With Actuator Traffic};
    \node[font=\footnotesize,align=center,anchor=north west] at ({rel axis cs:0.01,1}) {Prompt \& Reliable\\\it Always / LRU};
  \end{groupplot}
  \path (group c1r2.south west) -- (group c2r2.south east) node[midway,below=0.5cm,anchor=north] {Duration [min]};
  \path (group c1r2.south west) -- (group c1r1.north west) node[midway,sloped,above=0.55cm,anchor=south] {Packet Rate [$\frac{\#}{min}$]};
\end{tikzpicture}
%
  \caption{Packet transmission rate per minute for requests and responses measured at the gateway.}%
  \label{fig:baseline}
\end{figure*}
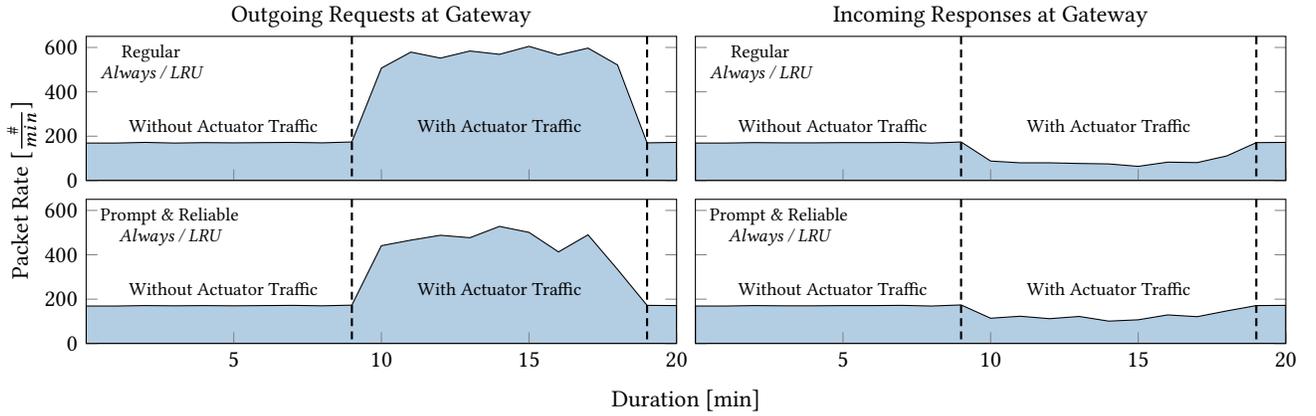
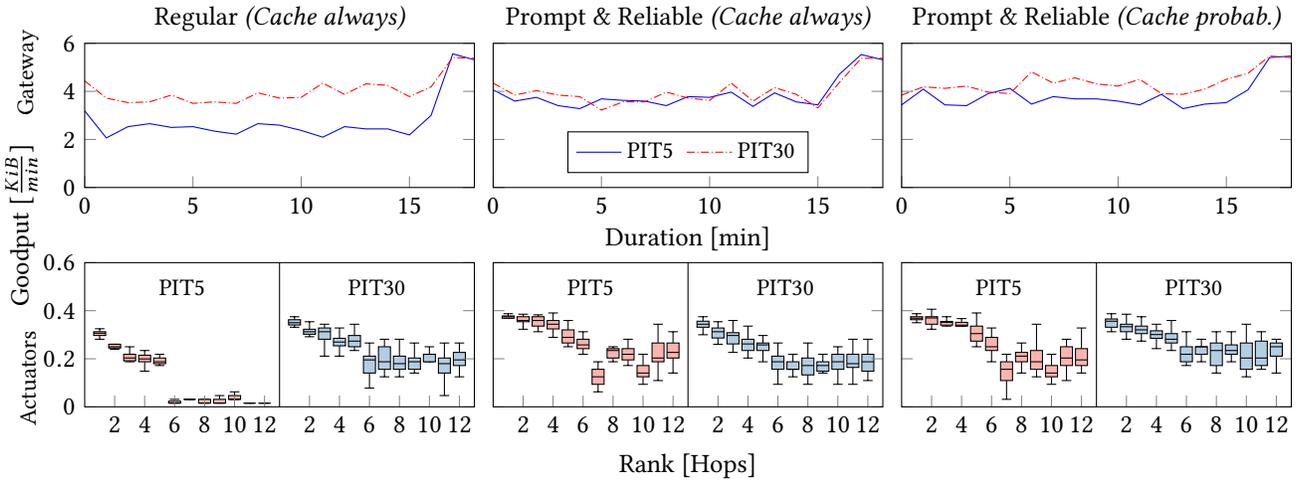
\begin{figure*}[t]
  \centering
  \tikzexternalenable
\tikzsetnextfilename{scenario1-goodputs-cs5}
\begin{tikzpicture}
  \begin{groupplot}[
    group style={
      group size=3 by 2,
      horizontal sep=0.25cm,
      vertical sep=1.00cm,
    },
    height=3.5cm,
    width=0.38\textwidth,
    no markers,
    boxplot/draw direction=y,
    xtick align=inside,
    xtick pos=left,
    axis on top,
    title style={anchor=north,yshift=0.4cm},
    legend style={at={(0.5,0.1)}, anchor=south,font=\small,legend columns=3,/tikz/every even column/.append style={column sep=0.1cm}},
    ]
    \nextgroupplot[ymin=0,ymax=6,xmin=0,xmax=18,align=center,title={Regular \it (Cache always)}]
    \addplot+[solid,blue] table [x index=0, y index=1] {data/sa_config1_pit5_CS5_goodputs.csv};
    \addplot+[densely dashdotted,red] table [x index=0, y index=1] {data/sa_config1_pit30_CS5_goodputs.csv};
    \nextgroupplot[yticklabels={},ymin=0,ymax=6,xmin=0,xmax=18,align=center,title={Prompt \& Reliable \it (Cache always)}]
    \addplot+[solid,blue] table [x index=0, y index=1] {data/sa_config3_pit5_CS5_goodputs.csv};
    \addplot+[densely dashdotted,red] table [x index=0, y index=1] {data/sa_config3_pit30_CS5_goodputs.csv};
    \legend{PIT5, PIT30};
    \nextgroupplot[yticklabels={},ymin=0,ymax=6,xmin=0,xmax=18,align=center,title={Prompt \& Reliable \it (Cache probab.)}]
    \addplot+[solid,blue] table [x index=0, y index=1] {data/sa_config15_pit5_CS5_goodputs.csv};
    \addplot+[densely dashdotted,red] table [x index=0, y index=1] {data/sa_config15_pit30_CS5_goodputs.csv};

    \nextgroupplot[ymin=0,ymax=0.6,xmin=0,xmax=26,xtick={2,4,6,8,10,12,15,17,19,21,23,25},xticklabels={2,4,6,8,10,12,2,4,6,8,10,12}]
    \node[font=\small,anchor=north] at ({rel axis cs:0.25,0.95}) {PIT5};
    \node[font=\small,anchor=north] at ({rel axis cs:0.75,0.95}) {PIT30};
    \draw[draw=black] ({axis cs:13,0}|-{rel axis cs:0,1}) -- ({axis cs:13,0}|-{rel axis cs:0,0});
    \foreach \i [count=\xi from 1] in {2,...,13} {
      \addplot[boxplot, boxplot/draw position=\xi, fill=Pastel1-A, draw=black] table[y index=\i] {data/sa_config1_pit5_CS5_goodputboxes.csv};
    }
    \foreach \i [count=\xi from 14]in {2,...,13} {
      \addplot[boxplot, boxplot/draw position=\xi, fill=Pastel1-B, draw=black] table[y index=\i] {data/sa_config1_pit30_CS5_goodputboxes.csv};
    }
    \nextgroupplot[ymin=0,ymax=0.6,xmin=0,xmax=26,yticklabels={},xtick={2,4,6,8,10,12,15,17,19,21,23,25},xticklabels={2,4,6,8,10,12,2,4,6,8,10,12}]
    \node[font=\small,anchor=north] at ({rel axis cs:0.25,0.95}) {PIT5};
    \node[font=\small,anchor=north] at ({rel axis cs:0.75,0.95}) {PIT30};
    \draw[draw=black] ({axis cs:13,0}|-{rel axis cs:0,1}) -- ({axis cs:13,0}|-{rel axis cs:0,0});
    \foreach \i [count=\xi from 1] in {2,...,13} {
      \addplot[boxplot, boxplot/draw position=\xi, fill=Pastel1-A, draw=black] table[y index=\i] {data/sa_config3_pit5_CS5_goodputboxes.csv};
    }
    \foreach \i [count=\xi from 14]in {2,...,13} {
      \addplot[boxplot, boxplot/draw position=\xi, fill=Pastel1-B, draw=black] table[y index=\i] {data/sa_config3_pit30_CS5_goodputboxes.csv};
    }
    \nextgroupplot[ymin=0,ymax=0.6,xmin=0,xmax=26,yticklabels={},xtick={2,4,6,8,10,12,15,17,19,21,23,25},xticklabels={2,4,6,8,10,12,2,4,6,8,10,12}]
    \node[font=\small,anchor=north] at ({rel axis cs:0.25,0.95}) {PIT5};
    \node[font=\small,anchor=north] at ({rel axis cs:0.75,0.95}) {PIT30};
    \draw[draw=black] ({axis cs:13,0}|-{rel axis cs:0,1}) -- ({axis cs:13,0}|-{rel axis cs:0,0});
    \foreach \i [count=\xi from 1] in {2,...,13} {
      \addplot[boxplot, boxplot/draw position=\xi, fill=Pastel1-A, draw=black] table[y index=\i] {data/sa_config15_pit5_CS5_goodputboxes.csv};
    }
    \foreach \i [count=\xi from 14]in {2,...,13} {
      \addplot[boxplot, boxplot/draw position=\xi, fill=Pastel1-B, draw=black] table[y index=\i] {data/sa_config15_pit30_CS5_goodputboxes.csv};
    }
  \end{groupplot}
  \path (group c1r2.south west) -- (group c1r1.north west) node[midway,sloped,above=0.5cm,anchor=south] {Goodput [$\frac{KiB}{min}$]};
  \path (group c1r2.south west) -- (group c3r2.south east) node[midway,below=0.5cm,anchor=north] {Rank [Hops]};
  \path (group c1r1.south west) -- (group c3r1.south east) node[midway,below=0.4cm,anchor=north] {Duration [min]};
  \path (group c1r1.south west) -- (group c1r1.north west) node[pos=0.80,above=0.75cm,anchor=base,sloped,anchor=center,font=\small] {Gateway};
  \path (group c1r2.south west) -- (group c1r2.north west) node[pos=0.20,above=0.75cm,anchor=base,sloped,anchor=center,font=\small] {Actuators};
\end{tikzpicture}
\tikzexternaldisable

%
  \caption{Goodput evolution for \textit{Scenario 1} with actuator and gateway traffic using a CS size of 5.}%
  \label{fig:scenario1_goodputs}
\end{figure*}

\paragraph{Network utilization}

In typical convergecast settings, a majority of the traffic traverses the gateway to reach remote endpoints.
Thus, a robust operation of the gateway node is crucial to ensure adequate performance of the entire IoT network.
Due to the increasing sequence numbers used in the naming scheme for \textit{Scenario 1}, virtually no response contributes to future in-network cache hits.
To gauge the network load on the gateway, we analyze the number of outgoing requests and incoming responses during a setup
in which actuator initiated traffic is added to the gateway initiated traffic after approximately eight minutes.
We first perform this experiment without any QoS features enabled, and then repeat it with the adjustment that
actuator initiated traffic is prioritized using the \textit{reliable} and \textit{prompt} service levels.
The PIT of the gateway is configured to  a maximum of 50 entries, while the remaining nodes have a PIT maximum of 5 entries.
The CS is limited to 5 entries for all nodes.

We observe in Figure~\ref{fig:baseline} that the gateway initiated traffic exhibits a steady request-response flow of about $180$ \textit{packets/min} 
for both requests and responses.
As soon as the actuators initiate their periodic requests (at minute eight), the network load at the gateway increases.
The number of requests spikes threefold due to network layer retransmissions, while the number of returning responses drops to half.
In contrast, the setup with prioritized actuator traffic clearly admits a reduced number of requests at the gateway while achieving a higher response rate.

These results reveal that prioritizing the actuator initiated traffic has a positive effect on the overall network load due to reduced retransmissions.

\paragraph{Goodputs}
Figure~\ref{fig:scenario1_goodputs} shows that while PIT sizes have a significant impact on the goodput at the gateway and at each actuator during normal operation,
this effect is reduced when QoS service classes are introduced.
When using service classes and the \emph{always} caching decision strategy,
 network members with small PIT sizes can reach a level of goodput that is comparable to those of the largest PIT size during normal operation.
With \emph{probabilistic} caching in operation, the network performance  increases even further for larger PITs, which support a higher number of concurrent flows that can leverage cache diversity better.  
This overall enhanced transport performance is caused by flows that complete with delay based on segment-wise available network capacities and retransmissions. The corresponding temporal effects can be observed from Figure~\ref{fig:scenario1_ttc_cs5}.

At actuator nodes far from the gateway,
our QoS mechanisms mitigate the effects of PIT exhaustion, which in normal operation leads to an abrupt 
collapse of the throughput at around a rank of $6$.
Our QoS mechanisms cause a smooth, gradual decline in performance instead.

\paragraph{Time to completion}
We can see from Figure~\ref{fig:scenario1_ttc_cs5} that content arrival times are significantly reduced for smaller PIT sizes when QoS service classes are introduced---for traffic at gateways and even more at the prioritized actuators.
Simultaneously, we see again a signature of enhanced traffic delivery for QoS-coordinated flows that shows doubled  success rates from 40\% to 80\%. 
It is worthwhile to (re-)observe that the sensor traffic also experiences improvements due to a more efficient balancing of resources---small PIT sizes in particular.

We now quantify the content arrival times separately for each quality dimension.
Figure~\ref{fig:scenario1_ttc_prompt_reliable} illustrates the nodal content arrival times with PIT and CS sizes  both set to 5.
In all three displayed configurations, completion time increases with the distance to the gateway.
While arrival times range from 5~ms to 350~ms for the majority of requests with the \textit{regular} and \textit{reliable} configurations,
the \textit{prompt} configuration yields noticeably faster delivery times for all nodes.
Especially nodes far away from the gateway experience a reduction of about 100~ms ($\approx 30$\%).

\subsubsection{Scenario 2: Sensing and Lighting Control}%
\begin{figure*}
  \centering
  \tikzexternalenable
\tikzsetnextfilename{scenario1-ttc-pit-cs5}
\begin{tikzpicture}
  \begin{groupplot}[
    group style={
      group size=3 by 2,
      horizontal sep=0.25cm,
      vertical sep=0.25cm,
    },
    height=3.5cm,
    width=0.38\textwidth,
    no markers,
    xtick align=inside,
    xtick pos=left,
    xtick={0,2,4,6,8},
    axis on top,
    title style = {yshift=0.4cm, anchor=north,align=center},
    legend style={at={(0.5,0.05)}, anchor=south,font=\small,legend columns=3,/tikz/every even column/.append style={column sep=0.1cm}},
    ]
    \nextgroupplot[ymin=0,ymax=1,xmin=0,xmax=10,title={Regular \it (Cache always)},xticklabels={},cycle list name=color]
    \addplot+[thick,dotted] table [x index=0, y index=1] {data/sa_config1_CS5_acdf.csv};
    \addplot+[thick,densely dotted] table [x index=0, y index=2] {data/sa_config1_CS5_acdf.csv};
    \addplot+[thick,dashed] table [x index=0, y index=3] {data/sa_config1_CS5_acdf.csv};
    \addplot+[thick,densely dashed] table [x index=0, y index=4] {data/sa_config1_CS5_acdf.csv};
    \addplot+[thick,densely dashdotted] table [x index=0, y index=5] {data/sa_config1_CS5_acdf.csv};
    \addplot+[thick,solid] table [x index=0, y index=6] {data/sa_config1_CS5_acdf.csv};
    \nextgroupplot[ymin=0,ymax=1,xmin=0,xmax=10,title={Prompt \& Reliable \it (Cache always)},xticklabels={},cycle list name=color,yticklabels={}]
    \addplot+[thick,dotted] table [x index=0, y index=1] {data/sa_config3_CS5_acdf.csv};
    \addplot+[thick,densely dotted] table [x index=0, y index=2] {data/sa_config3_CS5_acdf.csv};
    \addplot+[thick,dashed] table [x index=0, y index=3] {data/sa_config3_CS5_acdf.csv};
    \addplot+[thick,densely dashed] table [x index=0, y index=4] {data/sa_config3_CS5_acdf.csv};
    \addplot+[thick,densely dashdotted] table [x index=0, y index=5] {data/sa_config3_CS5_acdf.csv};
    \addplot+[thick,solid] table [x index=0, y index=6] {data/sa_config3_CS5_acdf.csv};
    \nextgroupplot[ymin=0,ymax=1,xmin=0,xmax=10,title={Prompt \& Reliable \it (Cache probab.)},xticklabels={},cycle list name=color,yticklabels={}]
    \addplot+[thick,dotted] table [x index=0, y index=1] {data/sa_config15_CS5_acdf.csv};
    \addplot+[thick,densely dotted] table [x index=0, y index=2] {data/sa_config15_CS5_acdf.csv};
    \addplot+[thick,dashed] table [x index=0, y index=3] {data/sa_config15_CS5_acdf.csv};
    \addplot+[thick,densely dashed] table [x index=0, y index=4] {data/sa_config15_CS5_acdf.csv};
    \addplot+[thick,densely dashdotted] table [x index=0, y index=5] {data/sa_config15_CS5_acdf.csv};
    \addplot+[thick,solid] table [x index=0, y index=6] {data/sa_config15_CS5_acdf.csv};
    \nextgroupplot[ymin=0,ymax=1,xmin=0,xmax=10,cycle list name=color]
    \addplot+[thick,dotted] table [x index=0, y index=1] {data/sa_config1_CS5_scdf.csv};
    \addplot+[thick,densely dotted] table [x index=0, y index=2] {data/sa_config1_CS5_scdf.csv};
    \addplot+[thick,dashed] table [x index=0, y index=3] {data/sa_config1_CS5_scdf.csv};
    \addplot+[thick,densely dashed] table [x index=0, y index=4] {data/sa_config1_CS5_scdf.csv};
    \addplot+[thick,densely dashdotted] table [x index=0, y index=5] {data/sa_config1_CS5_scdf.csv};
    \addplot+[thick,solid] table [x index=0, y index=6] {data/sa_config1_CS5_scdf.csv};
    \nextgroupplot[ymin=0,ymax=1,xmin=0,xmax=10,yticklabels={},cycle list name=color]
    \addplot+[thick,dotted] table [x index=0, y index=1] {data/sa_config3_CS5_scdf.csv};
    \addplot+[thick,densely dotted] table [x index=0, y index=2] {data/sa_config3_CS5_scdf.csv};
    \addplot+[thick,dashed] table [x index=0, y index=3] {data/sa_config3_CS5_scdf.csv};
    \addplot+[thick,densely dashed] table [x index=0, y index=4] {data/sa_config3_CS5_scdf.csv};
    \addplot+[thick,densely dashdotted] table [x index=0, y index=5] {data/sa_config3_CS5_scdf.csv};
    \addplot+[thick,solid] table [x index=0, y index=6] {data/sa_config3_CS5_scdf.csv};
    \legend{PIT5, PIT10, PIT15, PIT20, PIT25, PIT30}
    \nextgroupplot[ymin=0,ymax=1,xmin=0,xmax=10,yticklabels={},cycle list name=color]
    \addplot+[thick,dotted] table [x index=0, y index=1] {data/sa_config15_CS5_scdf.csv};
    \addplot+[thick,densely dotted] table [x index=0, y index=2] {data/sa_config15_CS5_scdf.csv};
    \addplot+[thick,dashed] table [x index=0, y index=3] {data/sa_config15_CS5_scdf.csv};
    \addplot+[thick,densely dashed] table [x index=0, y index=4] {data/sa_config15_CS5_scdf.csv};
    \addplot+[thick,densely dashdotted] table [x index=0, y index=5] {data/sa_config15_CS5_scdf.csv};
    \addplot+[thick,solid] table [x index=0, y index=6] {data/sa_config15_CS5_scdf.csv};
  \end{groupplot}
  \path (group c1r2.south west) -- (group c1r1.north west) node[midway,sloped,above=0.60cm,anchor=south] {CDF};
  \path (group c1r1.south west) -- (group c1r1.north west) node[pos=0.60,above=0.75cm,anchor=base,sloped,align=center,font=\small] {Actuators};
  \path (group c1r2.south west) -- (group c1r2.north west) node[pos=0.40,above=0.75cm,anchor=base,sloped,align=center,font=\small] {Gateway};
  \path (group c1r2.south west) -- (group c3r2.south east) node[midway,below=0.5cm,anchor=north] {Time to Completion [s]};
\end{tikzpicture}
\tikzexternaldisable

%
  \caption{Time to completion for \textit{Scenario 1} with actuator and gateway traffic using a CS size of 5.}%
  \label{fig:scenario1_ttc_cs5}
\end{figure*}
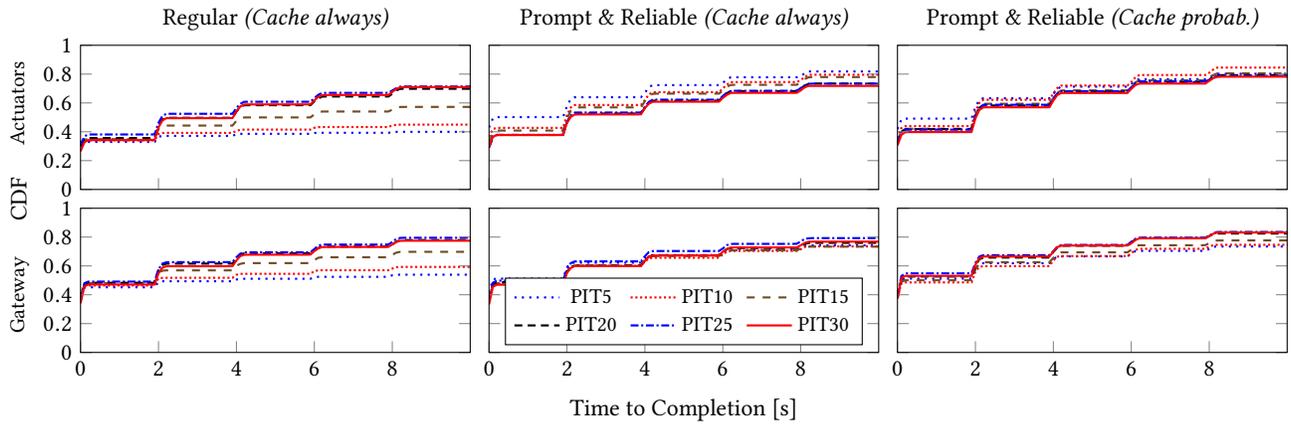
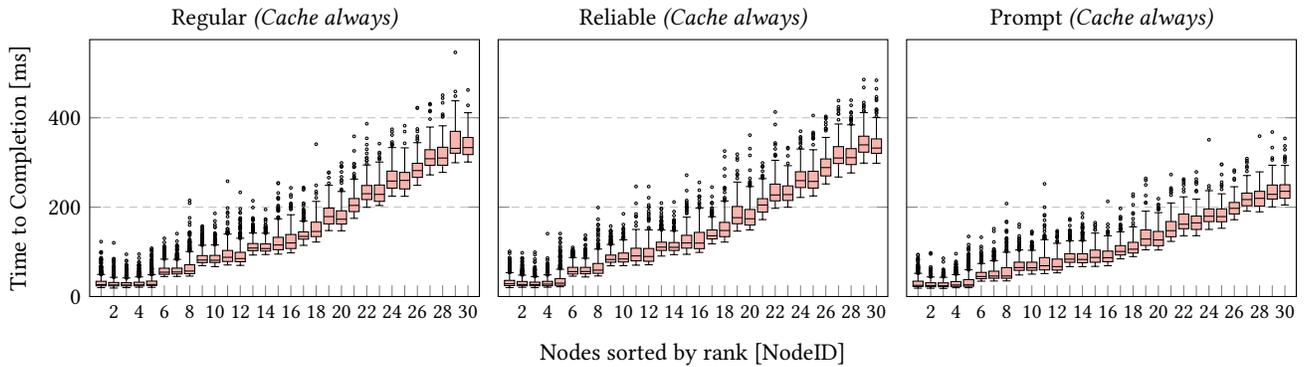
\begin{figure*}
  \centering
  \tikzexternalenable
\tikzsetnextfilename{scenario1-ttc-prompt-reliable}
\begin{tikzpicture}
  \begin{groupplot}[
    group style={
      group size=3 by 1,
      horizontal sep=0.25cm,
      vertical sep=0.4cm,
    },
    height=5cm,
    width=0.38\textwidth,
    /pgfplots/boxplot/whisker range={1.5},
    /pgfplots/boxplot/every median/.style={solid,black},
    boxplot/draw direction=y,
    xtick={1,...,30},
    xmin=0.10, xmax=30.90,
    ymin=0,ymax=575,
    xticklabel style = {font=\small,yshift=0pt},
    xticklabel={\pgfmathparse{mod(\ticknum+1,2)==0?int(\ticknum+1):}\pgfmathresult},
    yticklabel style = {xshift=0pt},
    ylabel style = {yshift=0pt},
    xtick align=inside,
    xtick pos=left,
    ymajorgrids,
    major grid style={line width=0.2pt, draw=gray!50, densely dashed},
    enlargelimits=false,
    title style={anchor=south,yshift=-0.2cm},
    legend style={at={(0.5,0.1)}, anchor=south,font=\small,legend columns=3,/tikz/every even column/.append style={column sep=0.1cm}},
    ]
    \nextgroupplot[align=center,title={Regular \it (Cache always)},ylabel={Time to Completion [ms]}]
    \foreach \i [count=\xi from 1] in {0,...,29} {
      \addplot[boxplot, boxplot/draw position=\xi, fill=Pastel1-A, draw=black,mark=*, mark size=0.5pt,every mark/.append style={fill=white,fill opacity=0,draw=black}] table[skip first n=1,y index=\i] {data/boxconfig1.csv};
    }
    \nextgroupplot[yticklabels={},align=center,title={Reliable \it (Cache always)}]
    \foreach \i [count=\xi from 1] in {0,...,29} {
      \addplot[boxplot, boxplot/draw position=\xi, fill=Pastel1-A, draw=black,mark=*, mark size=0.5pt,every mark/.append style={fill=white,fill opacity=0,draw=black}] table[skip first n=1,y index=\i] {data/boxconfig17.csv};
    }
    \nextgroupplot[yticklabels={},align=center,title={Prompt \it (Cache always)}]
    \foreach \i [count=\xi from 1] in {0,...,29} {
      \addplot[boxplot, boxplot/draw position=\xi, fill=Pastel1-A, draw=black,mark=*, mark size=0.5pt,every mark/.append style={fill=white,fill opacity=0,draw=black}] table[skip first n=1,y index=\i] {data/boxconfig18.csv};
    }
  \end{groupplot}
  \path (group c1r1.south west) -- (group c3r1.south east) node[midway,below=0.5cm,anchor=north] {Nodes sorted by rank [NodeID]};
\end{tikzpicture}
\tikzexternaldisable

%
  \caption{Time to completion per actuator and quality dimension in \textit{Scenario 1} using PIT and CS sizes of 5.}%
  \label{fig:scenario1_ttc_prompt_reliable}
\end{figure*}

In this scenario, prioritized actuator traffic flows from the gateway to multiple destinations and thus benefits from on-path caching. Accordingly, we expect the network performance to improve over that for Scenario~1.

\paragraph{Success rates}

The overall success values presented  in the second row of Figure~\ref{fig:scenario1and2_successes} confirm these expectations. Success in content delivery nicely approaches 100\% in most QoS settings.  
Even for the constrained number of 5 PIT entries, increasing the CS sizes from 5 to 30 suffices to turn traffic from QoS classes into a  reliable service.
While we do see some failures at higher ranks for the small CS size of 5 similar to  \emph{Scenario 1},
increasing the CS capacity to 10 is already sufficient to attain success rates above 80\%.
The most striking contrast is found in comparison to the results of regular NDN operation,
where even with a maximum CS size of 30 the failure rate stays at 30--40\% at higher ranks. Regular NDN traffic apparently profits less from cacheable content. This is mainly due to PIT decorrelation, which breaks content flows. 

\begin{figure*}
  \centering
  \tikzexternalenable
\tikzsetnextfilename{scenario2-ttc}
\begin{tikzpicture}
  \begin{groupplot}[
    group style={
      group size=3 by 2,
      horizontal sep=0.25cm,
      vertical sep=0.25cm,
    },
    height=3.5cm,
    width=0.38\textwidth,
    no markers,
    xtick align=inside,
    xtick pos=left,
    xtick={0,2,4,6,8},
    axis on top,
    title style = {yshift=0.4cm, anchor=north,align=center},
    legend style={at={(0.5,0.05)}, anchor=south,font=\small,legend columns=3,/tikz/every even column/.append style={column sep=0.1cm}},
    ]
    \nextgroupplot[ymin=0,ymax=1,xmin=0,xmax=10,title={Regular \it (Cache always)},xticklabels={},cycle list name=color]
    \addplot+[thick,dotted] table [x index=0, y index=1] {data/sa_config7_pit5_acdf.csv};
    \addplot+[thick,densely dotted] table [x index=0, y index=2] {data/sa_config7_pit5_acdf.csv};
    \addplot+[thick,dashed] table [x index=0, y index=3] {data/sa_config7_pit5_acdf.csv};
    \addplot+[thick,densely dashed] table [x index=0, y index=4] {data/sa_config7_pit5_acdf.csv};
    \addplot+[thick,densely dashdotted] table [x index=0, y index=5] {data/sa_config7_pit5_acdf.csv};
    \addplot+[thick,solid] table [x index=0, y index=6] {data/sa_config7_pit5_acdf.csv};

    \nextgroupplot[ymin=0,ymax=1,xmin=0,xmax=10,yticklabels={},title={Prompt \& Reliable \it (Cache always)},xticklabels={},cycle list name=color]
    \addplot+[thick,dotted] table [x index=0, y index=1] {data/sa_config8_pit5_acdf.csv};
    \addplot+[thick,densely dotted] table [x index=0, y index=2] {data/sa_config8_pit5_acdf.csv};
    \addplot+[thick,dashed] table [x index=0, y index=3] {data/sa_config8_pit5_acdf.csv};
    \addplot+[thick,densely dashed] table [x index=0, y index=4] {data/sa_config8_pit5_acdf.csv};
    \addplot+[thick,densely dashdotted] table [x index=0, y index=5] {data/sa_config8_pit5_acdf.csv};
    \addplot+[thick,solid] table [x index=0, y index=6] {data/sa_config8_pit5_acdf.csv};

    \nextgroupplot[ymin=0,ymax=1,xmin=0,xmax=10,yticklabels={},title={Prompt \& Reliable \it (Cache probab.)},xticklabels={},cycle list name=color]
    \addplot+[thick,dotted] table [x index=0, y index=1] {data/sa_config9_pit5_acdf.csv};
    \addplot+[thick,densely dotted] table [x index=0, y index=2] {data/sa_config9_pit5_acdf.csv};
    \addplot+[thick,dashed] table [x index=0, y index=3] {data/sa_config9_pit5_acdf.csv};
    \addplot+[thick,densely dashed] table [x index=0, y index=4] {data/sa_config9_pit5_acdf.csv};
    \addplot+[thick,densely dashdotted] table [x index=0, y index=5] {data/sa_config9_pit5_acdf.csv};
    \addplot+[thick,solid] table [x index=0, y index=6] {data/sa_config9_pit5_acdf.csv};

    \nextgroupplot[ymin=0,ymax=1,xmin=0,xmax=10,cycle list name=color]
    \addplot+[thick,dotted] table [x index=0, y index=1] {data/sa_config7_pit5_scdf.csv};
    \addplot+[thick,densely dotted] table [x index=0, y index=2] {data/sa_config7_pit5_scdf.csv};
    \addplot+[thick,dashed] table [x index=0, y index=3] {data/sa_config7_pit5_scdf.csv};
    \addplot+[thick,densely dashed] table [x index=0, y index=4] {data/sa_config7_pit5_scdf.csv};
    \addplot+[thick,densely dashdotted] table [x index=0, y index=5] {data/sa_config7_pit5_scdf.csv};
    \addplot+[thick,solid] table [x index=0, y index=6] {data/sa_config7_pit5_scdf.csv};

    \nextgroupplot[ymin=0,ymax=1,xmin=0,xmax=10,yticklabels={},cycle list name=color]
    \addplot+[thick,dotted] table [x index=0, y index=1] {data/sa_config8_pit5_scdf.csv};
    \addplot+[thick,densely dotted] table [x index=0, y index=2] {data/sa_config8_pit5_scdf.csv};
    \addplot+[thick,dashed] table [x index=0, y index=3] {data/sa_config8_pit5_scdf.csv};
    \addplot+[thick,densely dashed] table [x index=0, y index=4] {data/sa_config8_pit5_scdf.csv};
    \addplot+[thick,densely dashdotted] table [x index=0, y index=5] {data/sa_config8_pit5_scdf.csv};
    \addplot+[thick,solid] table [x index=0, y index=6] {data/sa_config8_pit5_scdf.csv};
    \legend{CS5, CS10, CS15, CS20, CS25, CS30}

    \nextgroupplot[ymin=0,ymax=1,xmin=0,xmax=10,yticklabels={},cycle list name=color]
    \addplot+[thick,dotted] table [x index=0, y index=1] {data/sa_config9_pit5_scdf.csv};
    \addplot+[thick,densely dotted] table [x index=0, y index=2] {data/sa_config9_pit5_scdf.csv};
    \addplot+[thick,dashed] table [x index=0, y index=3] {data/sa_config9_pit5_scdf.csv};
    \addplot+[thick,densely dashed] table [x index=0, y index=4] {data/sa_config9_pit5_scdf.csv};
    \addplot+[thick,densely dashdotted] table [x index=0, y index=5] {data/sa_config9_pit5_scdf.csv};
    \addplot+[thick,solid] table [x index=0, y index=6] {data/sa_config9_pit5_scdf.csv};

  \end{groupplot}
  \path (group c1r2.south west) -- (group c1r1.north west) node[midway,sloped,above=0.60cm,anchor=south] {CDF};
  \path (group c1r1.south west) -- (group c1r1.north west) node[pos=0.60,above=0.75cm,anchor=base,sloped,align=center,font=\small] {Actuators};
  \path (group c1r2.south west) -- (group c1r2.north west) node[pos=0.40,above=0.75cm,anchor=base,sloped,align=center,font=\small] {Gateway};
  \path (group c1r2.south west) -- (group c3r2.south east) node[midway,below=0.5cm,anchor=north] {Time to Completion [s]};

\end{tikzpicture}
\tikzexternaldisable

%
  \caption{Time to completion for \textit{Scenario 2} with actuator and gateway traffic using a PIT size of 5.}%
  \label{fig:scenario2_ttc}
\end{figure*}
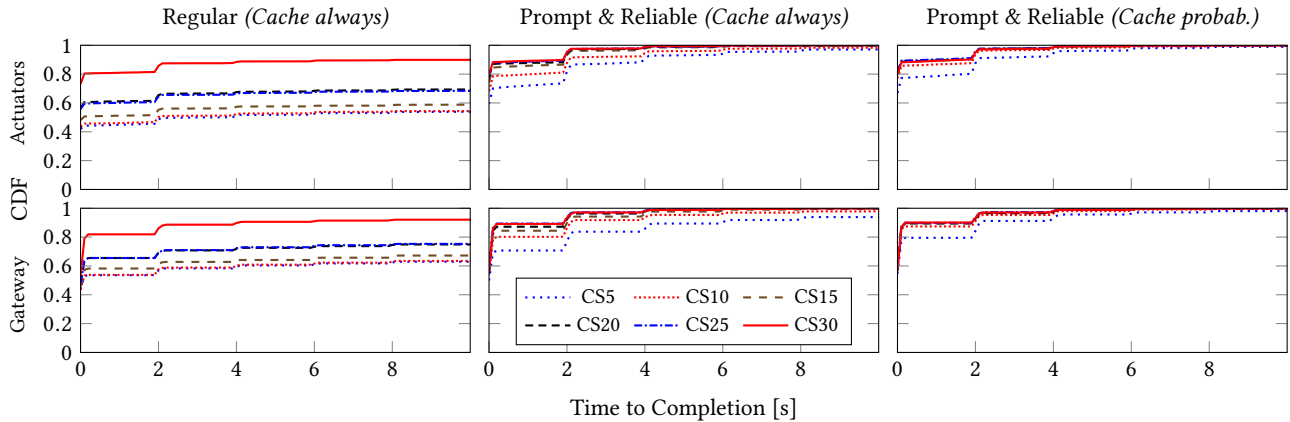

\paragraph{Time to completion}
 QoS service classes have a significant impact on the content arrival times for both the actuators and the gateway, as was already observed for \emph{Scenario~1}. Figure~\ref{fig:scenario2_ttc} reflects the same qualitative picture for \emph{Scenario~2}, but at  significantly reduced probabilities of retransmission. The latter is due to actuator traffic that is coordinated in QoS classes and coherently serviced from caches---a large reduction in overall network load. Results slightly improve for enhanced cache diversity in probabilistic caching, with 90\% of the packets arriving promptly ($< 100$~ms) at cache sizes of at least 10 packets.
Similar to \emph{Scenario~1}, CS sizes become less relevant in the presence of QoS marking, since prioritized traffic arrives quickly at its destination and remains unaffected by regular cache replacement. 

While we observe that increased CS sizes contribute to reduced completion times and improved success rates,
we also notice that even a CS size of 30 does not suffice for the \textit{regular} configuration.
On the other hand, the configurations that use QoS service levels display close to 100\% success rate,
even with severely limited CS sizes,
and about 70\% of all requests for each traffic type complete in less than 100~ms.

\paragraph{Cache hits}
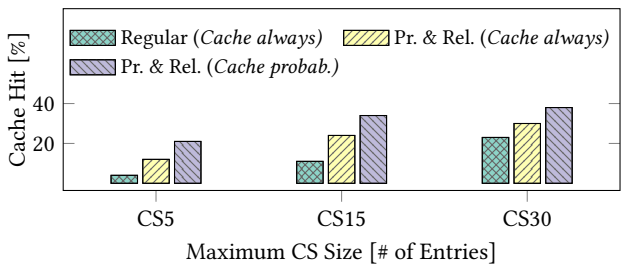
\begin{figure}
  \centering
  \begin{tikzpicture}
  \begin{axis}[
    width=1.06\columnwidth,
    height=4cm,
    enlargelimits,
    ybar,
    ymax=80,
    ytick={20,40},
    xtick=data,
    symbolic x coords={CS5,CS15,CS30},
    xtick pos=left,
    ylabel={Cache Hit [\%]},
    ylabel style={yshift=-5pt},
    xlabel={Maximum CS Size [\# of Entries]},
    enlarge x limits=0.25,
    legend cell align=left,
    area legend,
    legend style={draw=none,legend columns=2,anchor=north,at={(0.5,0.92)},/tikz/every even column/.append style={column sep=0pt},/tikz/every odd column/.append style={column sep=0pt},inner sep=0pt,outer sep=0pt,font=\small},
    ]
    \addplot[fill=Set3-A,postaction={pattern=crosshatch,pattern color=black!60}] coordinates {(CS5,4) (CS15,11) (CS30,23)};
    \addplot[fill=Set3-B,postaction={pattern=north east lines,pattern color=black!60}] coordinates {(CS5,12) (CS15,24) (CS30,30)};
    \addplot[fill=Set3-C,postaction={pattern=north west lines,pattern color=black!60}] coordinates {(CS5,21) (CS15,34) (CS30,38)};
    \legend{Regular (\it Cache always), Pr. \& Rel. (\it Cache always), Pr. \& Rel. (\it Cache probab.)};
  \end{axis}
\end{tikzpicture}%
  \caption{Cache hit for \textit{Scenario 2} and a PIT size of 5.}%
  \label{fig:scenario2_cachehits}
\end{figure}

Analysing the cache efficiencies supports these observations. 
Figure~\ref{fig:scenario2_cachehits} displays the relative in-network cache hits for actuator traffic in setups with varying CS sizes.
While the regular NDN operation yields a marginally improving cache hit ratio for increasing CS sizes,
both QoS enabled setups exhibit a noticeable enhancement.
This improved cache efficacy is caused by the privileged cache resource utilization for data of the \textit{reliable} actuator traffic,
whereas data of the gateway traffic is more likely to be evicted.
Another expected observation is that
\emph{probabilistic} caching further improves the cache hit ratio thanks to its increased CS diversity.


\section{Conclusions and Outlook}\label{sec:c+o}

We  presented and analyzed  QoS extensions to NDN that are suitable for constrained devices. Starting from a name-oriented flow classification scheme, we introduced the two service dimensions \emph{prompt} and \emph{reliable} network forwarding. Strategies were defined that not only foster local, isolated resource allocations, but take into account coordinative effects between different internal resources of a node, as well as correlations between nodes. Here we exploited the rich set of forwarding and caching options that NDN includes.\@

We were able to validate our approach in real-world experiments on a large testbed using a realistic multi-hop wireless setup. Moreover, we learned that QoS management in NDN is not confined to simple resource trading, but can lead to a global enhancement of network performance by optimizing the interplay between various resource consumptions. In particular, we found evidence that \one coordination of PIT and CS has a prevailing  effect on the overall performance of the networked system, and \two incorporation of Interests in  QoS treatment is vital to cater for resource coordination.     

In future work, we will enhance and optimize our service elements and implementations. Due to the constrained nature of our experimental platform, several resource decisions had to remain less elaborate than desired. In particular, the forwarding queues on RIOT had to stay limited due to memory constraints in this challenged regime. We further want to explore mechanisms to securely administer and distribute QoS settings at runtime.
Since in our experiment setups the high priority flows did not dominate the network, we plan to investigate the effects of our proposed QoS mechanisms in overbooked network settings that are threatened by service starvation.
An experimental deployment in our industrial environment is on schedule as well.


\subsection*{A Note on Reproducibility}
We fully support reproducible research~\cite{acmrep,swgsc-terrc-17} and perform all our experiments using open source software and an open access testbed.
Code and documentation will be available on Github at \url{https://github.com/5G-I3/ACM-ICN-2019-QOS}.
\subsection*{Acknowledgements}
This work was inspired by many fruitful discussions in the IRTF ICN research group, as well as by industrial deployment demands. Notably, we want to thank Dave Oran, who not only stimulated major design thinking, but also helped as a thoughtful shepherd to improve this paper. 
It was supported in part by the German Federal Ministry for Education and Research (BMBF) within the projects {\em I3 -- Information Centric Networking for the Industrial Internet}, {\em RAPstore -- RIOT App Store}, and the Hamburg {\em ahoi.digital} initiative with {\em SANE}.


\balance
\bibliographystyle{ACM-Reference-Format}
\bibliography{./local,own,rfcs,ids,ngi,iot,internet,layer2,meta}

\end{document}